\documentclass[10pt,conference]{IEEEtran}

\usepackage{cite}
\usepackage{amsmath,amssymb,amsfonts}   % amssymb is now safe (no newtxmath clash)
\usepackage{algorithmic}
\usepackage{graphicx}
\graphicspath{{./}{Figures/}}
\usepackage{textcomp}
\usepackage{xcolor}
\usepackage[hyphens]{url}
\usepackage{fancyhdr}
\usepackage[bookmarks=true,breaklinks=true,letterpaper=true,colorlinks,citecolor=blue,linkcolor=blue,urlcolor=blue]{hyperref}
\usepackage{booktabs}
\usepackage[normalem]{ulem}
\usepackage{multirow}
\usepackage{caption}
\usepackage{subcaption}
\usepackage{enumitem}
\usepackage{tikz}
\usetikzlibrary{decorations.pathreplacing, positioning, arrows.meta, calc}

\title{VIPER: Architecture-Aware Performance Modeling for Processing-in-Memory Design-Space Exploration}
\newcommand\hpcaauthors{
Haoran Geng$^{\dagger,*}$,
Tomas Sousa Pereira$^{\dagger,*}$,
Xiaoyang Lu$^{\ddagger}$,
Xian-He Sun$^{\ddagger}$,
Michael Niemier$^{\dagger}$,
and X. Sharon Hu$^{\dagger}$
}

\newcommand\hpcaaffiliation{
$^{\dagger}$University of Notre Dame, Notre Dame, IN, USA \\
$^{\ddagger}$Illinois Institute of Technology, Chicago, IL, USA \\
$^{*}$Equal contribution
}

\newcommand\hpcaemail{
\{hgeng, tsousape, mniemier, shu\}@nd.edu,
\{xlu40, sun\}@illinoistech.edu
}

\definecolor{deepgreen}{RGB}{34,139,34}
\definecolor{deepyellow}{RGB}{218,165,32}
\definecolor{viperYellowBg}{HTML}{FFF2CC}
\definecolor{viperYellowBorder}{HTML}{D6B656}
\definecolor{viperPurpleBg}{HTML}{E1D5E7}
\definecolor{viperPurpleBorder}{HTML}{9673A6}

\author{
  \IEEEauthorblockN{\hpcaauthors{}}
  \IEEEauthorblockA{
    \hpcaaffiliation{} \\
    \hpcaemail{}
  }
}

\begin{document}
\maketitle
\thispagestyle{empty}
\pagestyle{empty}

% Retained for compatibility with the paper source.
\newcommand{\hpcaheight}{0mm}

\begin{abstract}

Processing-in-Memory (PIM) promises to reduce data movement overhead by executing computation in or near memory, but its realized application speedup remains highly design-dependent. Non-offloadable host execution, host-PIM transfers, limited PIM capacity, and device programming latency can limit end-to-end speedup, making fast early-stage design-space exploration (DSE) essential. However, existing PIM evaluation methods remain limited: circuit- and device-level tools cannot capture these end-to-end PIM performance factors, while cycle-accurate simulation is too slow for iterative DSE.
To address this gap, we present VIPER, a unified, lightweight, and architecture-aware performance evaluation framework for PIM DSE.
VIPER profiles host execution once and combines the measured host behavior with a PIM-aware analytical engine that sweeps PIM-side parameters across candidate designs. It supports both Processing Near Memory (PNM) and Processing Using Memory (PUM) under task-offloading and data-triggered execution by capturing host-PIM transfer, array access, in-memory computation, device programming latency, and capacity-induced partitioning, providing rapid architecture-aware performance estimates for iterative DSE without repeated cycle-accurate simulation.
We validate VIPER against a commercial UPMEM system and more than 400 cycle-accurate gem5 configurations. VIPER predicts the UPMEM offloading decision and break-even region a priori, and, with a refined transfer model, captures the measured peak-and-rolloff behavior with 12\% mean speedup error across the DPU sweep (6\% up to the 256-DPU peak). Against gem5, VIPER achieves less than 10\% error while reducing evaluation time from hours to under one minute. Case studies of UPMEM, ReRAM/FeFET crossbars, and IMCRYPTO show that architecture-aware DSE  reveals key performance trade-offs that device-level evaluation misses.

% We then use VIPER to study UPMEM, ReRAM/FeFET crossbars, and IMCRYPTO, demonstrating that architecture-aware DSE across diverse PIM designs reveals key performance trade-offs for better utilizing the performance potential of PIM.
\end{abstract}

\section{Introduction}
\label{sec:introduction}

 Before a Processing-in-Memory (PIM) design is committed to silicon, the architect must determine whether its
system-level benefits justify its capacity, technology, and processing resources.
How much on-chip capacity should be provisioned, and what does falling
short cost? Which memory technology should host the computation, when
NVM programming is $10$--$50\times$ slower than
reading~\cite{XX_reram_write, ni2019fefet_tcam}? How many processing
units pay for themselves before host-side data movement erases the
gains? These are \emph{architecture-level} questions, must be
answered at the \emph{earliest} design stage, when the only artifacts
that exist are device parameters and a workload.

Our case
studies on real PIM designs (Section~\ref{sec:Result}) show that early
design decisions produce cliffs, not slopes: a crossbar array with
\emph{almost} enough capacity for its workload performs strictly worse
than one four times smaller, and missing the residency threshold costs
two orders of magnitude of speedup; a commercial near-memory system
benefits from its first few hundred processing units, then beyond the sweet spot, each additional unit reduces end-to-end speedup; and
for the same in-memory AES engine design, changing only the memory
technology that hosts its arrays decides viability: CMOS and FeFET keep the encryption overhead acceptable, while slower-programming RRAM and PCM do not. None of these outcomes is visible at the
circuit or device level, where PIM proposals are typically evaluated.
They arise from \emph{cross-layer} design parameter interactions such as, the architecture-level data transfer between host and PIM, the device-physics programming latency of NVM cells, capacity-driven partitioning of the working set, and non-offloadable host execution. Such interactions can only be studied  when the PIM device is
placed inside a full memory hierarchy and driven by a real
workload~\cite{gomez2021benchmarking, ghose2019workload, mutlu2020modern,
mannocci2023emerging, wan2022nature_rram, yin2024fefet_tcam}. Investigating these interactions essentially requires answer the following question: \textbf{what
design parameters determine whether a PIM design delivers
real performance benefit, and how can a designer quantify them before
committing to silicon?}

\begin{figure}[t]
\centering
\resizebox{1.00\columnwidth}{!}{\includegraphics{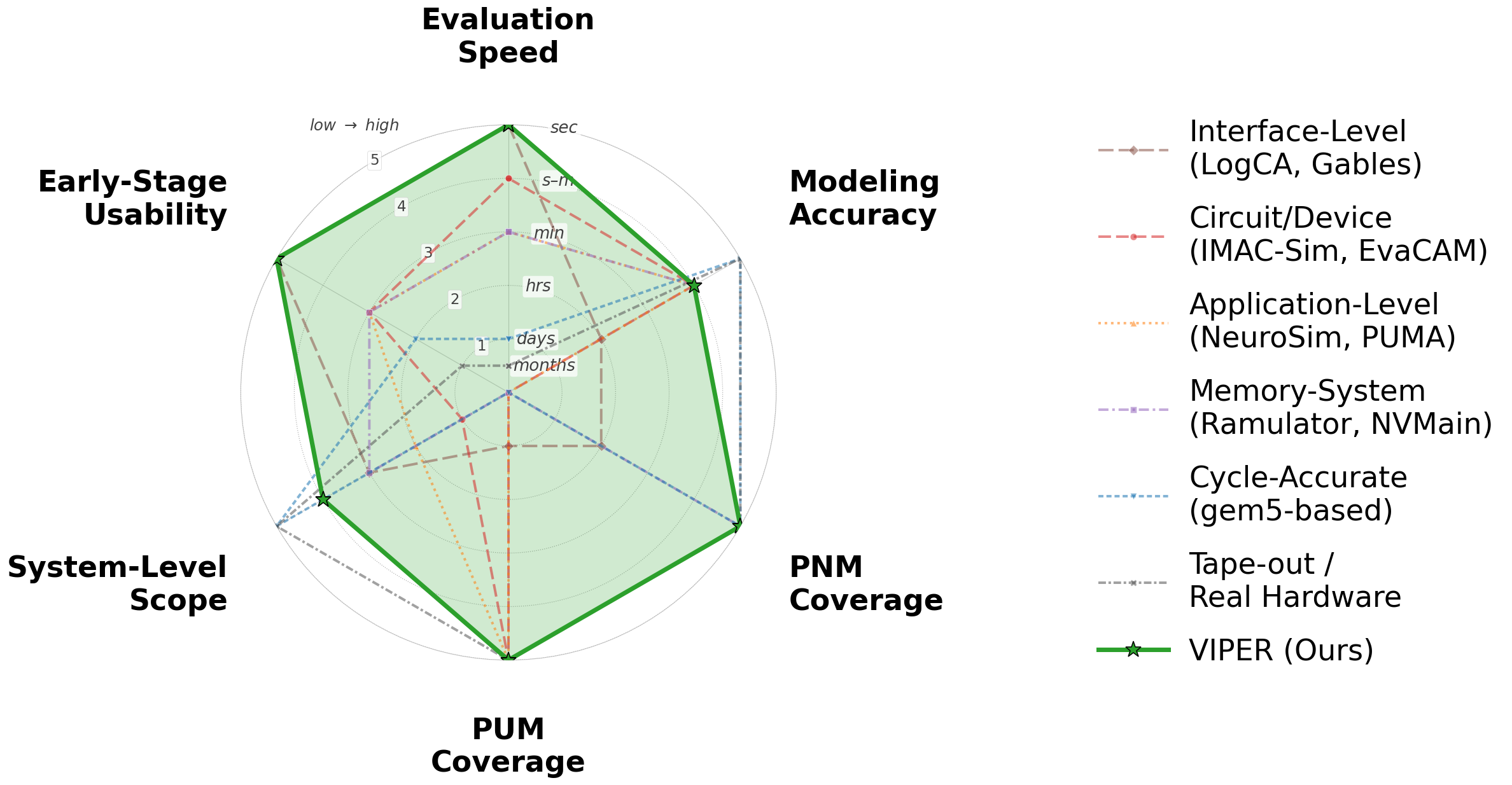}}
\vspace*{-5mm}
\caption{PIM evaluation tools across six dimensions. VIPER
pairs architecture-level modeling and PNM/PUM support with
evaluation in seconds, within 10\% of gem5.}
\label{fig:viper_poisition}
\vspace*{-5mm}
\end{figure}

Given the huge design space, answering this question requires fast early-stage design-space
exploration~(DSE) across both PIM paradigms: Processing Near
Memory~(PNM), which places processing elements adjacent to memory
arrays~\cite{devaux2019upmem, lee2021hbmpim, kim2022aquabolt,
ahn2015scalable, farmahini2015nda, boroumand2018google}, and Processing
Using Memory~(PUM), which computes in situ within the memory cells
themselves~\cite{seshadri2017ambit, seshadri2013rowclone,
hajinazar2021simdram, shafiee2016isaac, chi2016prime,
wan2022nature_rram, yin2022cosime, imcrypto, shared_pim}. As
Figure~\ref{fig:viper_poisition} illustrates, existing tool families
each sacrifice a dimension a PIM designer cannot do without.
Interface-level models such as LogCA and Gables~\cite{logca, gables}
evaluate in seconds, but reduce the host and memory system to a few
interface costs, missing in-memory computation, capacity partitioning,
and programming latency. Circuit- and device-level tools such as
IMAC-Sim and Eva-CAM~\cite{amin2022imacsim, liu2022evacam} are
accurate at the array boundary but model neither the workload nor its
interaction with the host and memory hierarchy; application-level simulators
such as NeuroSim, MNSIM, and CAMASim~\cite{chen2018neurosim,
ankit2019puma, zhu2023mnsim, wang2024pimsim, camasim_nd} are tied to
narrow application domains; memory-system simulators such as Ramulator,
DRAMsim3, and Sim2PIM~\cite{kim2015ramulator, poremba2012nvmain,
santos2022sim2pim, li2020dramsim3, kwon2021pimsimulator} detail the
memory but omit host execution and barely support PUM; and
cycle-accurate full-system simulators such as gem5-X, gem5-SALAM, and
PIMSim~\cite{qureshi2019gem5x, christ2023pimsys, lima2019pimgem5,
xu2019pimsim, ceary2020gem5salam, shao2016codesigning} model the whole
system but need 6--30 hours per configuration and heavy engineering per
design, making iterative exploration of hundreds of points impractical.
No existing tool provides fast, accurate, architecture-level modeling
for PIM DSE across both PNM and PUM.

We present VIPER,\footnote{VIPER is open source and available at
\url{https://github.com/Notre-Dame-HW-SW-Codesign-Lab/VIPER}.}, an analytical PIM DSE framework that fills this gap.
VIPER's key insight is to model PIM as a component of the memory
hierarchy rather than as a conventional host-attached accelerator, so
every PIM cost, from device physics to host interaction, enters the
memory access path directly. Building on the
classical Average Memory Access Time (AMAT) formulation, VIPER derives
a cross-layer PIM performance model with explicit terms for host--PIM
data transfer, data movement between memory arrays and PIM processing
units, NVM cell programming latency, and capacity-driven partitioning.
This single formulation spans PNM and PUM under both
task-offloading and data-triggered execution, and evaluates a
design point in seconds, letting a designer sweep hundreds of
configurations before any hardware or simulator exists.

In summary, this paper makes the following contributions:

\begin{itemize}

\item \textbf{A cross-layer analytical performance model for PIM.}
    We derive a analytical model that embeds PIM in the memory
    access path, capturing device-, architecture-, and execution-level
    latency for PNM and PUM under task-offloading and data-triggered
    execution, including data transfer overhead, array access latency,
    device programming overhead, and the partitioning overhead when the
    working set exceeds PIM capacity, providing a unified analytical
    foundation for DSE.

\item \textbf{An open-source tool for PIM DSE.} 
   We build the above cross-layer model into VIPER, which profiles the
host application once to fix all CPU-side parameters and then allows
the designer to explore PIM designs through dedicated task-offloading
and data-triggering flows, each exposing the design parameters
relevant to its execution model. This removes simulation from the
DSE loop: VIPER evaluates in seconds what requires 6--30 hours per
configuration in gem5. The VIPER framework, including source code
and evaluation scripts, is publicly available at
\url{https://github.com/Notre-Dame-HW-SW-Codesign-Lab/VIPER}.
\item \textbf{Validated performance estimation on commercial PIM hardware and cycle-accurate simulation.} Across 400+ gem5 runs spanning diverse PIM
    configurations, benchmarks, and CPU types, VIPER achieves below
    10\% mean estimation error, with 97\% of task-offloading
    predictions falling within a $\pm 10\%$ envelope. On a
    2{,}560-DPU UPMEM server~\cite{devaux2019upmem}, the first
    commercial PNM system, VIPER's a-priori prediction of a Llama2
    matmul offload, built entirely from host profiling and published
    UPMEM parameters, correctly anticipated the measured
    Amdahl ceiling, break-even DPU range, and offload decision before
    a single line of DPU code had been ported.

\item \textbf{Design-space insights on real PIM designs.} We apply VIPER to crossbar arrays on two beyond-CMOS devices (ReRAM, FeFET) and an in-memory AES accelerator to identify pre-silicon
limits: end-to-end speedup is limited by the work that stays on the
    CPU and by CPU--PIM data transfer, not by how fast the PIM device
    computes; an array slightly too small for its workload performs
    worse than a far smaller one, because the workload must be
    repeatedly split and reloaded; and the same PIM design can be
    fast enough in one memory technology yet too slow in another.

\end{itemize}

\section{Background}
\label{sec:background}

\subsection{Processing-in-Memory}

\begin{figure}[t]
\centering
\resizebox{\columnwidth}{!}{%
\begin{tikzpicture}[
    block/.style={draw, thick, rounded corners=2pt, minimum height=0.85cm,
                  text centered, font=\normalsize},
    memcell/.style={draw, thick, fill=blue!12, minimum width=0.7cm,
                    minimum height=0.7cm, font=\normalsize},
    pumcell/.style={draw, thick, fill=purple!18, minimum width=0.7cm,
                    minimum height=0.7cm, font=\normalsize, text centered},
    pe/.style={draw, thick, fill=orange!30, rounded corners=2pt,
               font=\normalsize, text centered},
    arr/.style={->, thick, >=stealth},
    darr/.style={<->, thick, >=stealth},
    lbl/.style={font=\normalsize, text centered},
]

%% ---- Top-level PIM banner ----
\node[draw, thick, rounded corners=3pt, fill=gray!12,
      minimum width=11.6cm, minimum height=0.7cm,
      font=\normalsize\bfseries] at (5.5, 5.8)
      {Processing-in-Memory (PIM)};
\draw[arr] (3.0, 5.45) -- (2.5, 5.1);
\draw[arr] (8.0, 5.45) -- (8.5, 5.1);

%% ===== (a) PNM side (left) =====
\node[font=\normalsize\bfseries] at (2.5, 4.9)
    {(a) Processing \emph{Near} Memory (PNM)};

% Host CPU
\node[block, fill=gray!18, minimum width=4.8cm] (cpu) at (2.5, 4.2) {Host CPU};

% Off-chip bus
\draw[darr, thick] (2.5, 3.77) --
    node[right, font=\normalsize]{off-chip bus} (2.5, 3.15);

% Memory chip boundary
\draw[thick, dashed, rounded corners=4pt] (-0.2, -0.15) rectangle (5.5, 3.05);
\node[font=\normalsize\bfseries, anchor=north west] at (-0.1, 3.05)
    {\textit{Memory Chip}};

% Processing element inside chip
\node[pe, minimum width=1.4cm, minimum height=1.4cm] (pe) at (0.8, 1.4)
    {\begin{tabular}{c}Processing\\Element\end{tabular}};

% Memory array (3x3 grid with visible gaps)
\foreach \r in {0,1,2} {
    \foreach \c in {0,1,2} {
        \node[memcell] at (3.25+\c*0.85, 0.55+\r*0.85) {};
    }
}
\node[lbl, text=blue!60!black] at (4.1, 0.0) {Memory Array};

% Data transfer arrow -- emphasized as bottleneck
\draw[darr, red!65!black, line width=1.4pt]
    (pe.east) -- node[above, font=\small,
    text=red!65!black, yshift=1pt]{\shortstack{\textbf{data}\\\textbf{mvmt.}}} (2.85, 1.4);

% Bottom brace annotation
\draw[thick, decorate, decoration={brace, amplitude=4pt, mirror}]
    (-0.2, -0.35) -- (5.5, -0.35)
    node[midway, below=5pt, font=\normalsize, align=center]
    {UPMEM, HBM-PIM\\(DRAM, HBM)};

%% ===== (b) PUM side (right) =====
\node[font=\normalsize\bfseries] at (8.5, 4.9)
    {(b) Processing \emph{Using} Memory (PUM)};

% Host CPU
\node[block, fill=gray!18, minimum width=4.8cm] (cpu2) at (8.5, 4.2) {Host CPU};

% Off-chip bus
\draw[darr, thick] (8.5, 3.77) --
    node[right, font=\normalsize]{off-chip bus} (8.5, 3.15);

% Memory chip boundary
\draw[thick, dashed, rounded corners=4pt] (5.8, -0.15) rectangle (11.2, 3.05);
\node[font=\normalsize\bfseries, anchor=north west] at (5.9, 3.05)
    {\textit{Memory Chip}};

% Memory array with in-cell compute (3x4 grid, each cell shows f)
\foreach \r in {0,1,2} {
    \foreach \c in {0,1,2,3} {
        \node[pumcell] at (7.15+\c*0.85, 0.55+\r*0.85)
            {\textcolor{purple!70!black}{$f$}};
    }
}
\node[lbl, text=purple!60!black] at (8.5, 0.0)
    {Memory Array \textit{(compute in-cell)}};

% Bottom brace annotation
\draw[thick, decorate, decoration={brace, amplitude=4pt, mirror}]
    (5.8, -0.35) -- (11.2, -0.35)
    node[midway, below=5pt, font=\normalsize, align=center]
    {Ambit, ISAAC, RowClone\\(RRAM, FeFET, PCM)};

\end{tikzpicture}%
}
\caption{PIM taxonomy. (a)~\textbf{PNM}: PEs adjacent to the array, with residual data movement. (b)~\textbf{PUM}: computation \emph{in situ} within cells, no internal transfers.}
\label{fig:pim_overview}
\end{figure}
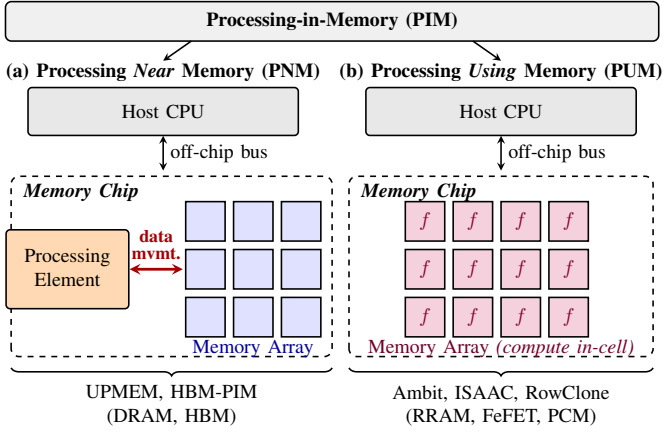

PIM integrates the computation directly into or near memory devices to mitigate the data movement bottleneck~\cite{mutlu2020modern, ghose2019workload}. PIM architectures are categorized into two main approaches: \textbf{PNM}, which places processing units adjacent to memory arrays, and \textbf{PUM}, which exploits the analog properties of memory cells to perform computation~\cite{gomez2021benchmarking}. Figure~\ref{fig:pim_overview} illustrates the architectural distinction between these two paradigms.

% \vspace*{-5mm}

\begin{figure*}[tb]
\centering
\includegraphics[width=\textwidth]{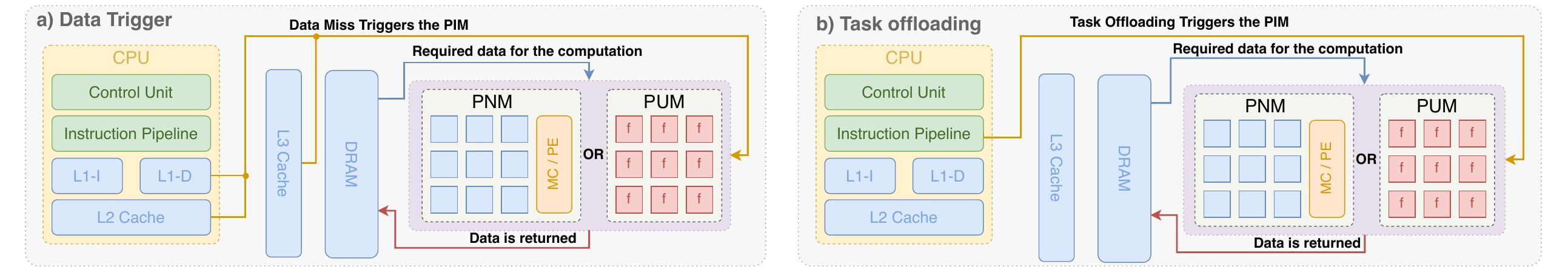}
\caption{PIM trigger models with different types of PIM.}
\vspace*{-3mm}
% \Description{PIM trigger models with different types of PIM.}
\label{fig:pim_triggering}

\end{figure*}

% \begin{table*}[!htbp]
% \centering
% \caption{Taxonomy of prior PIM triggering models: task offloading versus data-triggering approaches.}
% \vspace*{-3mm}
% \label{tab:pim_triggering}
% \small
% \begin{tabular}{lcccp{9cm}}
% \toprule
% \textbf{Triggering Model} & \textbf{Task Size} & \textbf{PIM Instr.?} & 
% \textbf{Talks to CPU?} & \textbf{Prior Work} \\
% \midrule
% Task Offloading & \textcolor{red}{Large} & \textcolor{green}{Yes} & \textcolor{green}{Yes} &
% SRAM~\cite{ppimce,aga2017computecaches,lin2022sramsurv},
% DRAM~\cite{shared_pim,pluto,seshadri2017ambit,seshadri2013rowclone,hajinazar2021simdram},
% NAND Flash~\cite{flash_cosmos,gao2021parabit},
% CAM~\cite{ni2019fefet_tcam,liu2023fefet_cam},
% RRAM/memristive~\cite{X_poly,shafiee2016isaac,chi2016prime,wan2022nature_rram,yan2023cim_safety,wu2023sot_cim},
% FeFETs~\cite{reis2018cim_fefet,kazemi2022fefet_hdc,yin2022cosime,yin2024fefet_tcam,huang2022cim_ferroelectrics,laguna2023fewshot},
% PCM \& MRAM~\cite{mannocci2023emerging,wu2023sot_cim},
% general~\cite{gomez2021benchmarking,mutlu2020modern,ahn2015scalable,farmahini2015nda} \\
% \midrule
% Data-Triggered & \textcolor{green}{Small} & \textcolor{red}{No} & \textcolor{red}{No} &
% Prefetching~\cite{prefetch1,prefetch2,prefetch3,prefetch4},
% compression~\cite{compress1,compress2,compress3,compress4,compress5,compress6},Secure Enclave~\cite{imcrypto,cosmos},
% general~\cite{tako}
% \\
% \bottomrule
% \end{tabular}
% \end{table*}

\textbf{Processing Near Memory (PNM).}
PNM integrates processing elements such as general-purpose
cores, SIMD units, and application-specific accelerators adjacent to or
within the peripheral circuitry of standard memory devices, including SRAM, DRAM,
and NAND Flash~\cite{gomez2021benchmarking, mutlu2020modern, ahn2015scalable,
farmahini2015nda}. 
Representative real-world implementations include the UPMEM architecture, which places 
general-purpose cores alongside each DRAM bank on a standard DDR4 DIMM~\cite{devaux2019upmem, 
gomez2021benchmarking}, and Samsung's HBM-PIM, which embeds FP16 SIMD units within 
an HBM2 stack~\cite{lee2021hbmpim, kim2022aquabolt}.

PNM offers greater flexibility than PUM, with processing elements ranging from programmable cores to specialized accelerators, enabling support for diverse workloads~\cite{mutlu2020modern, asifuzzaman2023survey}. However, PNM does not eliminate data movement, as transfers between memory arrays and processing units remain necessary~\cite{gomez2021benchmarking}. Consequently, PNM is best suited for memory-bound, data-parallel workloads with low arithmetic intensity and limited data reuse~\cite{gomez2021benchmarking, boroumand2018google, lee2021hbmpim}.

\textbf{Processing Using Memory (PUM)} 
PUM exploits the existing memory architecture and the 
operational principles of memory cells and circuitry to perform operations directly 
within each memory chip at low cost~\cite{mutlu2020modern, seshadri2017ambit, 
seshadri2013rowclone}. Prior works have proposed PUM mechanisms by modifying the basic circuitry 
of SRAM~\cite{imcrypto, ppimce, aga2017computecaches, lin2022sramsurv}, 
DRAM~\cite{shared_pim, pluto, seshadri2017ambit, seshadri2013rowclone, 
hajinazar2021simdram}, NAND Flash~\cite{flash_cosmos, gao2021parabit}, and CAM~\cite{ni2019fefet_tcam, 
liu2023fefet_cam}. Beyond conventional memory technologies, PUM has attracted 
significant attention for its potential with emerging non-volatile memory (NVM) 
devices, which offer unique physical properties that naturally support in-situ 
computation. These include RRAM/memristive 
devices~\cite{X_poly, shafiee2016isaac, chi2016prime, wan2022nature_rram, 
yan2023cim_safety, wu2023sot_cim}, ferroelectric FETs 
(FeFETs)~\cite{reis2018cim_fefet, kazemi2022fefet_hdc, yin2022cosime, 
yin2024fefet_tcam, huang2022cim_ferroelectrics, laguna2023fewshot}, and phase-change 
memory (PCM) and MRAM~\cite{mannocci2023emerging, wu2023sot_cim}.

PUM offers a fundamental advantage over PNM by performing computation entirely within memory arrays, eliminating data movement overheads that PNM cannot fully avoid. Emerging NVM-based PUM designs, such as RRAM and FeFET crossbars, can achieve exceptional energy efficiency and throughput for targeted workloads, sometimes surpassing custom ASICs~\cite{wan2022nature_rram, kazemi2022fefet_hdc, chi2016prime}. However, PUM remains constrained by narrow, operation-specific compute models, interference with conventional memory accesses when deployed as main memory, and high write costs in emerging NVM devices~\cite{ni2019fefet_tcam, liu2023fefet_cam, shafiee2016isaac, shared_pim, mannocci2023emerging, wan2022nature_rram, yan2023cim_safety}.

% In this paper, we consider both PNM and PUM in a unified framework that explicitly 
% accounts for the full system overhead of each paradigm: the data transfer latency 
% between processing units and memory in PNM, and the programming latency of emerging 
% non-volatile memory cells in PUM. We instantiate our model using real PNM and PUM 
% hardware as case studies in Section~\ref{sec:Evaluation}, grounding our analysis in the 
% behavior of real, deployed silicon rather than idealized assumptions.

\subsection{PIM Triggering Model}

We categorize PIM architectures into two fundamental models based on how the host core initiates computation. 
Figure~\ref{fig:pim_triggering} illustrates the architectural difference between the two models.

\textbf{Task Offloading} is the dominant PIM paradigm, where the host CPU explicitly delegates computation to the PIM unit via dedicated ISA instructions~\cite{ppimce, aga2017computecaches, lin2022sramsurv, shared_pim, pluto, seshadri2017ambit, seshadri2013rowclone, hajinazar2021simdram, flash_cosmos, gao2021parabit, ni2019fefet_tcam, liu2023fefet_cam, X_poly, shafiee2016isaac, chi2016prime, wan2022nature_rram, yan2023cim_safety, wu2023sot_cim, reis2018cim_fefet, kazemi2022fefet_hdc, yin2022cosime, yin2024fefet_tcam, huang2022cim_ferroelectrics, laguna2023fewshot, mannocci2023emerging, gomez2021benchmarking, mutlu2020modern, ahn2015scalable, farmahini2015nda}. PIM executes large tasks directly where data resides, reducing data movement but requiring new ISA extensions and explicit CPU synchronization. The offload is synchronous: after dispatching a task, the CPU waits at the synchronization point until the PIM device completes.

\textbf{Data-Triggering} activates PIM automatically in response to hardware memory events such as cache misses or DRAM accesses, requiring no explicit CPU instruction~\cite{prefetch1, prefetch2, prefetch3, prefetch4, compress1, compress2, compress3, compress4, compress5, compress6, imcrypto, cosmos, tako}. This model is primarily used in narrow domains such as compression, prefetching, and secure memory operations. Because PIM execution lies on the critical memory access path, it is suitable only for small, low-latency tasks.

\section{Motivations}
\label{sec:Motivation}

PIM designers face two recurring challenges that are difficult to address without a fast, architecture-aware modeling tool. First, cross-layer costs such as architecture-level data transfer overhead, device-physics programming latency, and capacity constraints are often invisible at the circuit and device level. Second, no existing tool is fast enough to support iterative DSE that 
captures these costs across both PNM and PUM paradigms.

\begin{figure}[t]
\centering
\resizebox{1.00\columnwidth}{!}{\includegraphics{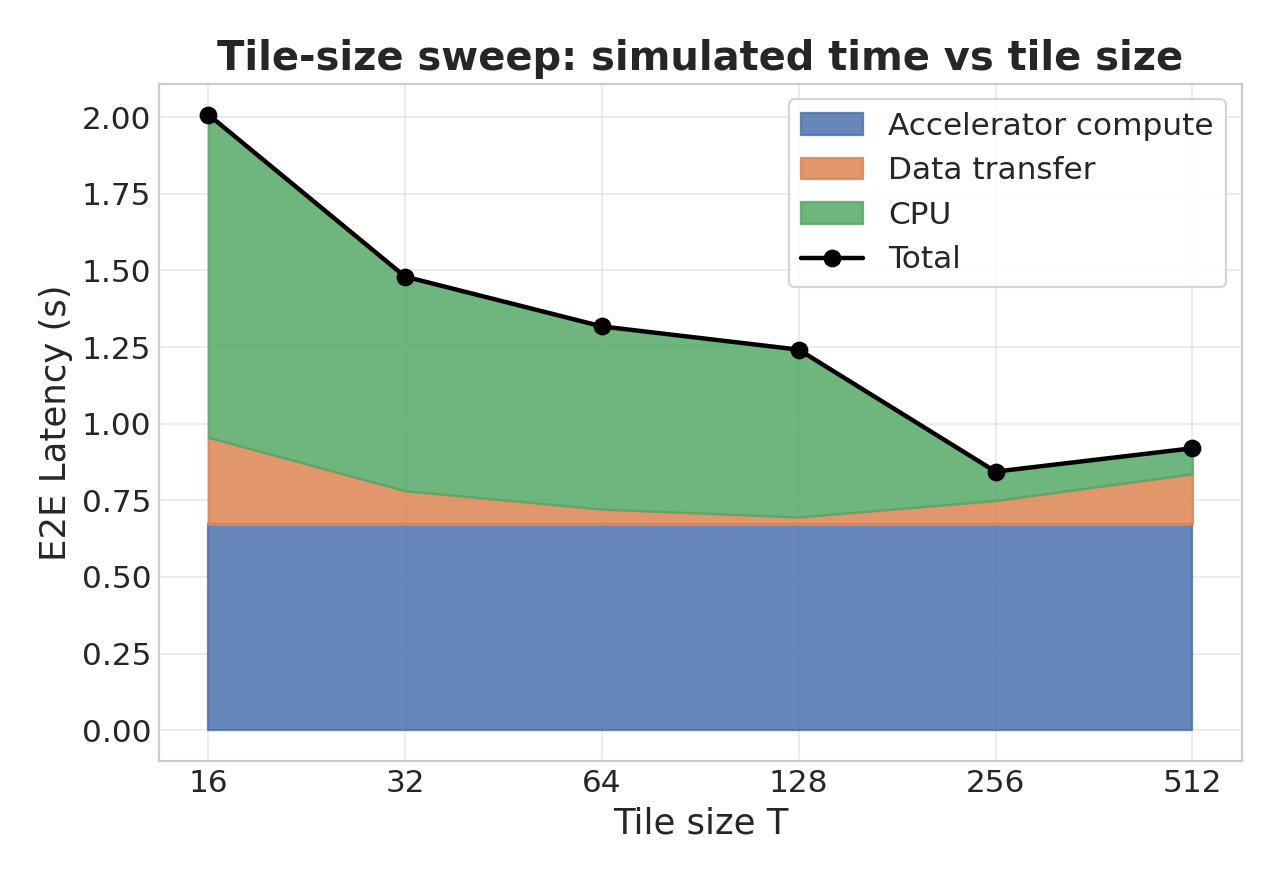}}
\vspace*{-8mm}
\caption{Tile-size sweep showing the breakdown of simulated time into accelerator compute, data transfer, and CPU components.}
\label{fig:tile_total_time}
\vspace*{-5mm}
\end{figure}

\subsection{Circuit- and Device-Level Evaluation Cannot Quantify End-to-End PIM Benefit}
PIM proposals are typically evaluated at the circuit and device
level, reporting latency, energy, and throughput at the array
boundary. End-to-end speedup, however, is determined by
cross-layer costs outside that boundary: host--PIM transfer,
array access overhead, device programming latency, capacity
constraints, and non-offloadable host execution. A device-level
metric alone cannot tell a designer whether a PIM design delivers
real benefit once integrated with a host and driven by a real
workload.

To illustrate, we sweep the tile size
$T \in \{16, 32, 64, 128, 256, 512\}$ for a fixed $N{=}512$ GEMM
offloaded to a SALAM-generated accelerator in
gem5~\cite{ceary2020gem5salam}, where the CPU stages
$T{\times}T$ sub-tiles into the accelerator scratchpad via DMA
and accumulates partial results back into DRAM. At the device
level, every configuration is identical: the same datapath
performs the same $N^3$ work. Yet
Figure~\ref{fig:tile_total_time} reveals a sweet spot at
$T{=}256$ (0.844\,s), with runtime rising at smaller $T$ due to
excessive CPU staging and at $T{=}512$ due to data-transfer cost
in the accelerator's dataflow pipeline. The breakdown explains
why: compute is constant, data transfer is U-shaped (bytes scale
as $N^3/T$, while large unrolled dataflow graphs lose pipeline
efficiency at $T{=}512$), and CPU staging time drops sharply as
dispatches fall from 64 to 8. The optimum is set entirely by
cross-layer costs that no array-boundary evaluation can expose.

Device programming latency introduces another cost, determined by device physics but paid at the architecture level. Writing an NVM cell requires a physical switching process, such as filament formation in RRAM, a phase transition in PCM, or ferroelectric domain switching in FeFET, which can take orders of magnitude longer than a CMOS write~\cite{mannocci2023emerging, wan2022nature_rram,yin2024fefet_tcam}. Cell-level characterization reports this latency, but not how often a system pays it: in a PUM deployment, every operand written into the array, and every reload forced by limited capacity, places this latency on the critical memory-access path. Slow programming can thus become a memory wall inside the PIM device itself, an effect that appears only when the device is evaluated within the memory hierarchy under a real workload.

\begin{center}
\setlength{\fboxrule}{1.5pt}
\fcolorbox{violet!70!black}{violet!15}{%
\parbox{0.95\linewidth}{%
\vspace{4pt}
\textcolor{violet!70!black}{%
\textbf{Motivation 1:} Early-stage PIM DSE must be
architecture-aware. Designers need to evaluate
host--memory--PIM integration factors, including host--PIM
transfer, array access overhead, device programming latency,
and capacity constraints, to determine whether a PIM design
can deliver end-to-end speedup.}
\vspace{2pt}
}}
\end{center}

\subsection{No Existing Tool Bridges the Gap}

A PIM designer navigating the existing tool landscape faces a fundamental trade-off at every level of abstraction. Fig.~\ref{fig:viper_poisition} summarizes each tool family across six dimensions: evaluation speed, modeling accuracy, PNM and PUM coverage, system-level scope, and early-stage usability.

% As we examine each category below, the same trade-off surfaces repeatedly: designers must choose between modeling accuracy, system visibility, architectural generality, and simulation speed, 
% but no existing tool delivers all four. This gap is what motivates VIPER.

Circuit and array level tools such as IMAC-Sim~\cite{amin2022imacsim} and EFSim~\cite{efsim2024} automatically generate SPICE netlists or Verilog circuits to evaluate accuracy, power, and latency at the device level, and EvaCAM~\cite{liu2022evacam} extends this to CAM-based designs. While these tools deliver high device-level accuracy, they are fundamentally scoped to the array itself, providing no visibility into host CPU interaction, data movement overhead, or capacity sufficiency. Application-level simulators such as NeuroSim~\cite{chen2018neurosim}, PUMA~\cite{ankit2019puma}, and CAMASim~\cite{camasim_nd} bridge the gap to workload behavior, but are scoped to narrow application domains and specific architectures, with simulation times ranging from minutes to hours per configuration. DRAM and memory-system simulators such as Ramulator~\cite{kim2015ramulator}, NVMain~\cite{poremba2012nvmain}, and Sim2PIM~\cite{santos2022sim2pim} bring the simulation closer to real system conditions but treat the host CPU as a black box and offer limited or no PUM support. gem5-based frameworks such as gem5-X~\cite{qureshi2019gem5x}, PIMSys~\cite{christ2023pimsys}, and gem5-Aladdin~\cite{shao2016codesigning} provide the most complete system-level view, but each configuration requires hours to days of simulation time, and extending these frameworks to support new PIM designs requires significant modification of the simulator source code.

\textbf{Existing tools do not combine fast architecture-aware modeling with support for both PNM and PUM.} 

\begin{center}
\setlength{\fboxrule}{1.5pt}
\fcolorbox{violet!70!black}{violet!15}{%
\parbox{0.95\linewidth}{%
\textcolor{violet!70!black}{%
\textbf{Motivation 2:} We aim to design a fast, accurate, 
architecture-aware modeling framework for PIM designers, one 
that captures full-system behavior across both PNM and 
PUM paradigms without sacrificing simulation speed.}
}}
\end{center}

\subsection{AMAT-Based Modeling Is All You Need}

% The survey above reveals that full-system simulators like gem5 are too slow for iterative PIM DSE, while lightweight analytical models sacrifice accuracy. This raises a natural question: \textbf{what is the right abstraction level for a practical PIM simulator?}

 The AMAT model provides a useful balance between accuracy and evaluation speed for early-stage PIM DSE. AMAT models the average time to complete a memory access as a weighted sum of hit latencies and miss penalties across the cache hierarchy. Unlike general performance models, AMAT is uniquely suited to PIM because PIM is fundamentally a memory device that sits in the memory hierarchy and whose cost is paid on every memory access that reaches it. Introducing PIM overhead directly into the AMAT model is therefore intuitive rather than approximate, naturally capturing near-memory processing cost, data transfer overhead, and capacity partitioning within the same familiar framework~\cite{ghose2019processing}. 

Analytical performance models have a long history of guiding
early design decisions at exactly this stage. Roofline
models~\cite{roofline} bound achievable throughput,
Gables~\cite{gables} extends them to heterogeneous SoCs,
and LogCA~\cite{logca} models accelerator offloading with
five parameters that expose break-even granularities and
speedup bounds before any hardware is built. VIPER follows
this lineage but targets what these models cannot express:
PIM is not an accelerator attached to the host, it is a
device inside the memory hierarchy. LogCA abstracts the host
and memory system into flat interface costs, and therefore
cannot capture data-triggered execution, capacity
partitioning, or device programming latency. Grounding the
model in AMAT rather than an interface-cost abstraction is
what makes these PIM-specific costs first-class terms.

AMAT models memory access time as a sequential stall cost and does not capture memory-level parallelism (MLP) or concurrent overlapping of PIM device costs. For early-stage PIM DSE, this is appropriate for three reasons. First, AMAT has been the standard foundation for memory hierarchy analysis precisely because sequential stall cost captures the dominant first-order effect on execution time, even under out-of-order execution where MLP exists. Second, at early PIM design stages, the memory scheduler, bank interleaving policy, and DMA overlap strategy are not yet fixed, so assuming sequential costs gives a conservative, architecture-independent baseline that a designer can refine as these decisions are made. Third, MLP and overlap depend primarily on the memory scheduler and system architecture. Our validation in Section~\ref{sec:val_data}  confirms that this assumption produces reliable trend prediction even under out-of-order execution.

\begin{center}
\setlength{\fboxrule}{1.5pt}
\fcolorbox{green!70!black}{green!15}{%
\parbox{0.95\linewidth}{%
\textcolor{green!70!black}{%
\textbf{Solution:} VIPER extends AMAT with PIM-specific timing and capacity terms to estimate architecture-level performance
without cycle-accurate simulation.
}
}}
\end{center}
\section{Architecture-aware Analytical Performance Model}
\label{sec:Model}

\begin{table}[t]
\centering
\caption{Summary of notation used in the analytical performance model.}
\vspace*{-3mm}
\label{tab:notation}
\small
\begin{tabular}{lp{6cm}}
\toprule
\textbf{Symbol} & \textbf{Description} \\
\midrule
\multicolumn{2}{l}{\textit{Execution Time}} \\
$T$                         & Total program execution time \\
$\text{CC}_{CPU}$                   & Clock cycle time on the CPU \\
$\text{CC}_{PIM}$        & Clock cycle time on the PIM device \\
\midrule
\multicolumn{2}{l}{\textit{Instructions}} \\
$I_{CPU}$                   & Number of instructions executed on the CPU \\
$I_{PIM}$                   & Number of instructions executed on the PIM device \\
$\text{CPI}_{CPU,\text{exec}}$ & Cycles per instruction on the CPU \\
$\text{CPI}_{PIM,\text{exec}}$ & Effective cycles per instruction on the PIM device \\

\midrule
\multicolumn{2}{l}{\textit{Memory Access}} \\
$m_{CPU}$                   & Average memory accesses per instruction on the CPU \\
$m_{PIM}$                   & Number of PIM invocations determined by capacity \\
$m_{L1}, m_{L2}, m_{L3}$   & Miss rates at L1, L2, and L3 cache levels \\

\midrule
\multicolumn{2}{l}{\textit{Latency}} \\
$t_{L1}, t_{L2}, t_{L3}$   & Hit latency at L1, L2, and L3 cache levels (cycles) \\
$t_{DRAM}$                  & DRAM access latency (cycles) \\
$t_{PIM,\text{mem}}$        & Effective PIM access latency (cycles)  \\
\midrule
\multicolumn{2}{l}{\textit{PIM Device}} \\
$t_{in}$                    & Data transfer time from host to PIM (s)  \\
$t_{comp}$                  & Computation time on the PIM device (s)  \\
$t_{array}$                 & Data movement from memory array to processing element (s)  \\
$t_{out}$                   & Data transfer time from PIM to host (s)  \\
$t_{prog}$                  & Cell programming latency for NVM-based PUM devices (s)  \\
$C_{PIM}$                   & PIM device capacity constraint \\
$D_{PIM}$                   & Data size of the offloaded or triggered PIM task \\
\midrule
\multicolumn{2}{l}{\textit{AMAT}} \\
$\text{AMAT}_{CPU}$         & Average memory access time for the CPU \\
$\text{AMAT}_{PIM}$         & Average memory access time for the PIM device \\
\bottomrule
\end{tabular}
\vspace*{-5mm}
\end{table}

In this section we present an AMAT-based analytical performance model for PIM architectures. Table~\ref{tab:notation} summarizes the notation used throughout this section, and the meaning of the notation will be discussed when they are first used.

\subsection{Task Offloading Model}
In the task offloading model, the host CPU explicitly delegates a 
portion of the program to the PIM device. The CPU memory access 
time follows the standard cache-hierarchy AMAT:
\begin{equation}
\label{eq:amat_cpu}
\text{AMAT}_{CPU} = t_{L1} + m_{L1} \left[ t_{L2} + m_{L2} \left( 
t_{L3} + m_{L3} \cdot t_{DRAM} \right) \right]
\end{equation}
The total execution time is the sum of CPU and PIM execution time:
\begin{align}
\label{eq:task_offload}
T &= I_{CPU} \times \left(\text{CPI}_{CPU,\text{exec}} + m_{CPU} 
\times \text{AMAT}_{CPU}\right) \times \text{CC}_{CPU} \notag \\
&+ I_{PIM} \times \left(\text{CPI}_{PIM,\text{exec}} + m_{PIM} 
\times t_{PIM,\text{mem}}\right) \times \text{CC}_{PIM}
\end{align}
where $t_{PIM,\text{mem}}$ captures the latency of a the PIM unit as a whole, and is expanded into PNM and PUM device parameters in 
Section~\ref{sec:pim_timing}.

\subsection{Data-Triggered Execution Model}
In the data-triggered model, PIM execution is activated automatically 
by a memory access event rather than an explicit CPU instruction. 
When a triggering event occurs, $t_{DRAM}$ in the standard AMAT is 
replaced by $t_{PIM,\text{mem}}$, giving the PIM-aware AMAT:
\begin{equation}
\label{eq:amat_pim}
\text{AMAT}_{PIM} = t_{L1} + m_{L1} \left[ t_{L2} + m_{L2} \left( 
t_{L3} + m_{L3} \cdot t_{PIM,\text{mem}} \right) \right]
\end{equation}

Accesses that hit in the cache hierarchy do not invoke the PIM and
retain their ordinary latencies; Equation~\ref{eq:amat_pim} assumes
every main-memory access triggers the PIM, which matches always-on
data-triggered designs such as in-memory encryption.
The total execution time then becomes:
\begin{equation}
\label{eq:data_trigger}
T = I_{CPU} \times \left(\text{CPI}_{CPU,\text{exec}} + m_{CPU} \times 
\text{AMAT}_{PIM}\right) \times \text{CC}_{CPU}
\end{equation}
The expansion of $t_{PIM,\text{mem}}$ into PNM and PUM device 
parameters is presented in Section~\ref{sec:pim_timing}.

\subsection{PIM Timing Model}
\label{sec:pim_timing}

We now derive the components of $t_{PIM,\text{mem}}$ for both triggering models across PNM and PUM paradigms. PNM pays for data movement from the memory array to the processing element; PUM eliminates this cost but may incur cell programming latency instead. When the working set exceeds the PIM unit's capacity, the task must be partitioned into multiple invocations. The capacity partitioning model uses a $\lceil D_{PIM}/C_{PIM} \rceil$ multiplier that captures the dominant first-order effect; bank-level parallelism and non-uniform data layout are second-order refinements that designers can incorporate by setting $C_{PIM}$ to the effective per-partition capacity of the target device.

\subsubsection{Task Offloading: PNM} 
In PNM task offloading, the host CPU explicitly offloads a computation to the PIM processing element (PE). Four costs are incurred: input data transfer from host to PIM ($t_{in}$), data movement from the memory array to the PE ($t_{array}$), output transfer back to host ($t_{out}$), and capacity partitioning overhead ($m_{PIM}$). The PIM memory access term, normalized to PIM clock cycles, is:

\begin{equation}
\label{eq:pnm_task_offload}
t_{PIM,\text{mem}} = \frac{t_{in} + t_{array} + t_{out}}{\text{CC}_{PIM}}
\end{equation}
\begin{equation}
\label{eq:m_pim}
m_{PIM} = \begin{cases} 
1 & \text{if } D_{PIM} \leq C_{PIM} \\[6pt]
\left\lceil \dfrac{D_{PIM}}{C_{PIM}} \right\rceil & \text{if } D_{PIM} > C_{PIM}
\end{cases}
\end{equation}

\subsubsection{Task Offloading: PUM} 
In PUM task offloading, computation occurs directly within the memory cells, eliminating $t_{array}$. Instead,  the model includes $t_{prog}$  before in-cell computation. This programming latency is inherent to NVM
technologies such as RRAM and FeFET and cannot be eliminated by architectural optimization alone.

\begin{equation}
t_{PIM,\text{mem}} = \frac{t_{in} + t_{prog} + t_{out}}{\text{CC}_{PIM}}
\end{equation} 

In both paradigms, $m_{PIM}$ follows the same capacity partitioning model as Equation~\ref{eq:m_pim}.

\subsubsection{Data Triggering: PNM}
In data triggering, the entire PIM execution cost is absorbed into the memory access latency. The cost additionally includes $t_{comp}$, since the computation sits directly on the critical memory access path. As in task offloading, the device times are normalized to clock cycles, here by the CPU clock $\text{CC}_{CPU}$, because $t_{PIM,\text{mem}}$ replaces $t_{DRAM}$ in the CPU-side AMAT (Equation~\ref{eq:amat_pim}). When $D_{PIM} > C_{PIM}$, the entire pipeline repeats for each partition, multiplying all cost terms:

\begin{equation}
t_{PIM,\text{mem}} = \left\lceil \frac{D_{PIM}}{C_{PIM}} \right\rceil \times
\frac{t_{in} + t_{array} + t_{comp} + t_{out}}{\text{CC}_{CPU}}
\end{equation}
where the ceiling factor reduces to one when the working set fits
($D_{PIM} \leq C_{PIM}$).

\subsubsection{Data Triggering: PUM}
In PUM data triggering, $t_{array}$ is eliminated since computation occurs in-cell, and $t_{prog}$ replaces it on the critical path:

\begin{equation}
t_{PIM,\text{mem}} = \left\lceil \frac{D_{PIM}}{C_{PIM}} \right\rceil \times
\frac{t_{in} + t_{prog} + t_{comp} + t_{out}}{\text{CC}_{CPU}}
\end{equation}

\section{VIPER: Design Space Exploration Framework}
\label{sec:Evaluation}

% VIPER translates the analytical model of Section~\ref{sec:Model} into a practical DSE tool.
% Table~\ref{tab:dse_summary} summarizes the framework. VIPER profiles the application once to fix CPU-side parameters, then sweeps designer-specified PIM parameters to minimize execution time. This separation means a designer can evaluate hundreds of PIM configurations in seconds without re-running the application or modifying any simulator source code.

VIPER turns the model of Section~\ref{sec:Model} into a DSE tool
(Table~\ref{tab:dse_summary}) by exploiting one structural property of PIM
execution: an application's host-side behavior, including its
instruction mix, cache behavior, and non-offloadable work, does
not depend on which PIM design is attached. CPU-side quantities
are therefore measured once and reused, while only the PIM-side
terms of Equations \ref{eq:task_offload} to \ref{eq:data_trigger} change per design point, removing simulation from the DSE loop entirely.

% ----------------------------------------------------------------
%  Compact DSE summary table
% ----------------------------------------------------------------
\begin{table}[tb]
\centering
\caption{VIPER design-space exploration summary.
\emph{Fixed} parameters are obtained once by profiling;
\emph{swept} parameters define the PIM design space.}
\label{tab:dse_summary}
\resizebox{\columnwidth}{!}{%
\begin{tabular}{@{}l l l@{}}
\toprule
& \textbf{Task Offloading} & \textbf{Data Triggering} \\
\midrule
\multicolumn{3}{@{}l}{\textbf{CPU-side (profiled via \texttt{perf}, fixed)}} \\
\midrule
Instructions        & $I_{CPU}$ (retired instructions)          & $I_{CPU}$ \\
Exec.\ CPI          & $\text{CPI}_{CPU,\text{exec}}$            & $\text{CPI}_{CPU,\text{exec}}$ \\
Memory access rate  & $m_{CPU}$ (L1 loads / instructions)       & $m_{CPU}$ \\
Cache hierarchy     & $\text{AMAT}_{CPU}$ ($m_{L1}, m_{L3}$,    & $m_{L1}, m_{L2}, m_{L3}$ \\
                    & \ \ $t_{L1}, t_{L3}, t_{DRAM}$)           & $t_{L1}, t_{L2}, t_{L3}$ \\
Clock period        & $\text{CC}_{CPU}$                                 & $\text{CC}_{CPU}$ \\
\midrule
\multicolumn{3}{@{}l}{\textbf{PIM-side (designer-swept)}} \\
\midrule
Instruction count   & $I_{PIM}$ (offloaded work)                & --- \\
Compute efficiency  & $\text{CPI}_{PIM,\text{exec}}$            & --- \\
PIM memory latency  & $t_{PIM,\text{mem}}$                      & $t_{PIM,\text{mem}}$ \\
                    & \ \ PNM: $(t_{in}\!+\!t_{array}\!+\!t_{out})/\text{CC}_{PIM}$ & \ \ (replaces $t_{DRAM}$ in AMAT) \\
                    & \ \ PUM: $(t_{in}\!+\!t_{prog}\!+\!t_{out})/\text{CC}_{PIM}$  & \\
Device capacity     & $C_{PIM}$ ($\rightarrow m_{PIM}$ via Eq.~\ref{eq:m_pim}) & $C_{PIM}$ ($\rightarrow$ partitioning) \\
Clock period        & $\text{CC}_{PIM}$                      & --- \\
\midrule
\multicolumn{3}{@{}l}{\textbf{DSE outputs}} \\
\midrule
Primary metric      & \multicolumn{2}{l}{End-to-end execution time $T$} \\
Derived insights    & \multicolumn{2}{l}{Speedup vs.\ baseline, break-even point (e.g.\ min.\ DPUs),} \\
                    & \multicolumn{2}{l}{Amdahl ceiling, sensitivity to $t_{prog}$/$t_{array}$/$C_{PIM}$,} \\
                    & \multicolumn{2}{l}{capacity wall threshold, technology trade-offs} \\
\bottomrule
\end{tabular}
}%
\vspace*{-8mm}
\end{table}

\subsection{Profiling Stage: Fixing the CPU Baseline}

Both execution models begin with the same profiling step.
VIPER uses Linux \texttt{perf}  to collect hardware-counter data, including a single instrumentation pass, the retired instruction count~($I_{CPU}$), total cycles, L1/LLC cache loads and misses.
From these counters VIPER derives the quantities listed in the \emph{CPU-side} rows of Table~\ref{tab:dse_summary}:

\begin{itemize}[nosep,leftmargin=*]
  \item \textbf{Execution-only CPI.}\;
    $\text{CPI}_{CPU,\text{exec}}$ is isolated by subtracting estimated memory-stall cycles from the measured CPI, using a hierarchical cache model with configurable penalties ($t_{L3}$ and $t_{DRAM}$; VIPER defaults are 10 and 100~cycles).
  \item \textbf{Memory access rate.}\;
    $m_{CPU}$ is the ratio of L1 data-cache loads to retired instructions.
  \item \textbf{AMAT.}\;
    $\text{AMAT}_{CPU}$ is computed from per-level miss rates ($m_{L1}, m_{L2}, m_{L3}$) and hit latencies ($t_{L1}, t_{L2}, t_{L3}, t_{DRAM}$) via Equation~\eqref{eq:amat_cpu}.
  \item \textbf{Clock period.}\;
    $\text{CC}_{CPU}$ is the ratio of wall-clock time to total cycles.
\end{itemize}

\noindent
Once measured, all CPU-side quantities are treated as fixed constants for the remainder of the exploration.

\subsection{Sweeping Stage: Exploring the PIM Design Space}

With the CPU baseline fixed, the designer sweeps the PIM-side parameters listed in Table~\ref{tab:dse_summary}.
The two execution models differ in which parameters are exposed and how PIM cost enters the total execution time; we describe each in turn.

\subsubsection{Task Offloading}
\label{sec:dse_task}

Figure~\ref{fig:viper_flow}(a) illustrates the task-offloading flow.
The fixed/swept partition corresponds directly to the two terms of Equation~\eqref{eq:task_offload}: the CPU term is held constant while the PIM term defines the design space (see Table~\ref{tab:dse_summary}).
Each swept parameter (Table~\ref{tab:dse_summary}) maps to a design decision: $I_{PIM}$ and $\text{CPI}_{PIM,\text{exec}}$ set how much work is offloaded and how efficiently it runs, $\text{CC}_{PIM}$ and $t_{PIM,\text{mem}}$ encode the device technology, and $C_{PIM}$ encodes capacity provisioning. The sweep therefore covers the design space itself rather than abstract model inputs. By sweeping them against the fixed baseline, VIPER identifies which changes yield the greatest reduction in end-to-end execution time and identifies break-even points and Amdahl ceilings that device-level models do not capture.

\subsubsection{Data Triggering}
\label{sec:dse_data}

\begin{figure}[tb]
\centering
% ---------- Shared TikZ styles ----------
\tikzset{
  panelmark/.style={font=\bfseries},
  cpubox/.style={
    draw=viperYellowBorder, line width=0.9pt, rounded corners=3pt,
    fill=viperYellowBg, fill opacity=0.7, draw opacity=0.9,
    minimum width=2.9cm, minimum height=0.55cm,
    inner sep=2.5pt, align=center, font=\footnotesize
  },
  pimbox/.style={
    draw=viperPurpleBorder, line width=0.9pt, rounded corners=3pt,
    fill=viperPurpleBg, fill opacity=0.7, draw opacity=0.9,
    minimum width=2.9cm, minimum height=0.55cm,
    inner sep=2.5pt, align=center, font=\footnotesize
  },
  wideyellow/.style={
    draw=viperYellowBorder, line width=0.9pt, rounded corners=3pt,
    fill=viperYellowBg, fill opacity=0.7, draw opacity=0.9,
    minimum width=3.2cm, minimum height=0.62cm,
    inner sep=3pt, align=center, font=\footnotesize
  },
  widepurple/.style={
    draw=viperPurpleBorder, line width=0.9pt, rounded corners=3pt,
    fill=viperPurpleBg, fill opacity=0.7, draw opacity=0.9,
    minimum width=3.2cm, minimum height=0.62cm,
    inner sep=3pt, align=center, font=\footnotesize
  },
  totalbox/.style={
    draw=black, line width=0.9pt, rounded corners=3pt,
    minimum width=1.0cm, minimum height=0.55cm,
    inner sep=2pt, align=center, font=\footnotesize
  },
  arr/.style={-{Stealth[length=4pt]}, line width=0.9pt},
  lbl/.style={font=\scriptsize, inner sep=1pt, fill=white, fill opacity=0.85, text opacity=1}
}
% ---------- Top panel ----------
\begin{tikzpicture}[x=1cm,y=1cm]
\node[panelmark] at (-0.35,0) {(a)};
\node[cpubox]     (app1) at (1.9, 0.0) {Normal CPU Application};
\node[cpubox]     (pro1) at (1.9,-1.0) {profile (e.g.\ \texttt{perf})};
\node[wideyellow] (cpu1) at (1.9,-2.15)
  {$\mathrm{CPI}_{CPU,\mathrm{exec}}$\\[-1pt]$+\,m_{CPU}\times\mathrm{AMAT}_{CPU}$};
\node[pimbox]     (des1) at (6.18, 0.0) {PIM designer};
\node[widepurple] (pim1) at (6.18,-2.15)
  {$\mathrm{CPI}_{PIM,\mathrm{exec}}$\\[-1pt]$+\,m_{PIM}\times t_{PIM,\mathrm{mem}}$};
\node[totalbox]   (tot1) at (4.1,-3.6) {$T$};
\draw[arr] (app1) -- node[right,lbl] {offload} (pro1);
\draw[arr] (pro1) -- node[right,lbl] {fixed} (cpu1);
\draw[arr] (des1) -- node[right,lbl] {swept} (pim1);
\draw[arr] (cpu1.south) |- ([xshift=-2pt]tot1.west)
  node[pos=0.32,right,lbl] {$I_{CPU}(\cdots)\text{CC}_{CPU}$};
\draw[arr] (pim1.south) |- ([xshift=2pt]tot1.east)
  node[pos=0.32,left,lbl] {$I_{PIM}(\cdots)t_{PIM,\mathrm{clk}}$};
\end{tikzpicture}

\vspace{2mm}

% ---------- Bottom panel ----------
\begin{tikzpicture}[x=1cm,y=1cm]
\node[panelmark] at (-0.35,0) {(b)};
\node[cpubox]     (app2) at (1.9, 0.0) {Normal CPU Application};
\node[cpubox]     (pro2) at (1.9,-1.0) {profile (e.g.\ \texttt{perf})};
\node[wideyellow] (cpu2) at (1.9,-2.15)
  {$m_{L1},m_{L2},m_{L3},$\\[-1pt]$t_{L1},t_{L2},t_{L3}$};
\node[pimbox]     (des2) at (6.18, 0.0) {PIM designer};
\node[widepurple] (pim2) at (6.18,-2.15)
  {$t_{PIM,\mathrm{mem}},\,C_{PIM}$};
\node[totalbox]   (tot2) at (4.1,-3.6) {$T$};
\draw[arr] (app2) -- node[right,lbl] {trigger} (pro2);
\draw[arr] (pro2) -- node[right,lbl] {fixed} (cpu2);
\draw[arr] (des2) -- node[right,lbl] {swept} (pim2);
\draw[arr] (cpu2.south) |- ([xshift=-2pt]tot2.west)
  node[pos=0.32,right,lbl] {$\mathrm{AMAT}_{CPU}$};
\draw[arr] (pim2.south) |- ([xshift=2pt]tot2.east)
  node[pos=0.32,left,lbl] {replace $t_{DRAM}$};
\end{tikzpicture}
\caption{VIPER flow for (a)~task offloading and (b)~data triggering.
 Yellow = profiled (fixed); purple = designer-swept.}
\vspace{-5mm}
\label{fig:viper_flow}
\end{figure}
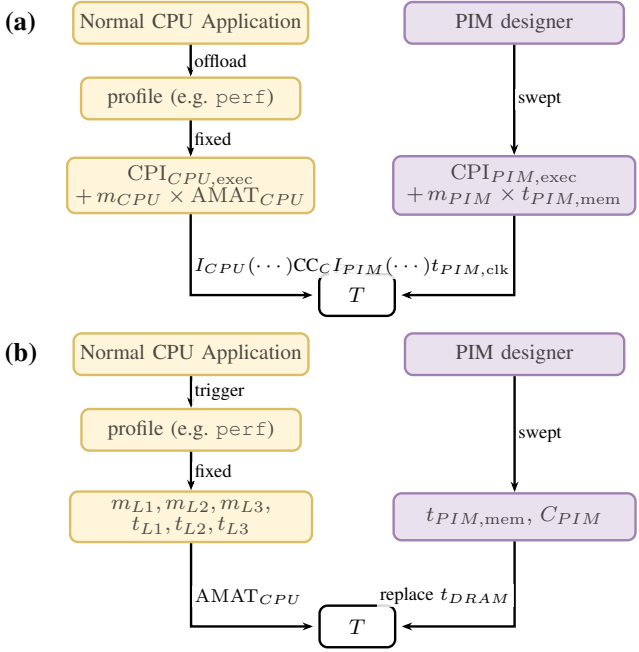

Figure~\ref{fig:viper_flow}(b) illustrates the data-triggering flow.
Here the CPU still executes the full program, but every memory access that reaches the PIM device pays $t_{PIM,\text{mem}}$ instead of $t_{DRAM}$ (Equation~\eqref{eq:amat_pim}).
Only two parameters are swept because a triggered design has no offloading decision: the trigger is architectural, so the entire design freedom is what the device costs per access ($t_{PIM,\text{mem}}$) and whether it avoids partitioning ($C_{PIM}$).
This two-dimensional sweep identifies the minimum capacity required to a avoid capacity-induced slowdown and the maximum $t_{PIM,\text{mem}}$ that still yields a net benefit over baseline DRAM latency.

% ---- Figure (unchanged from original) ----
\section{VIPER Validation}
\label{sec:validation}

We validate VIPER at two tiers. In this section, we compare against
cycle-accurate gem5-SALAM~\cite{ceary2020gem5salam} across the full
design space, verifying the task offloading and data triggering
models separately; in Section~\ref{sec:upmem_case}, we validate
VIPER's predictions against measurements on real PIM silicon. Simulation is the reference for breadth: the design points in Tables~\ref{tab:task_off_configs}
and~\ref{tab:val_configs} span SRAM, DRAM, HBM, NVM, and
storage-class PIM at arbitrary latencies and capacities, a space no
silicon implementation covers, and gem5 is the standard pre-silicon
reference, itself extensively validated against real
hardware~\cite{qureshi2019gem5x, gem5_accuracy}. 

% To further reduce
% dependence on absolute simulator fidelity, the data triggering
% evaluation compares normalized performance trends rather than raw
% execution times (Section~\ref{sec:val_data}).

% In this section, we validate VIPER against gem5-SALAM
% \cite{ceary2020gem5salam} to quantify how closely our analytical
% model predicts real execution behavior, verifying the task
% offloading and data triggering models separately. We validate
% against cycle-accurate simulation rather than silicon for a
% structural reason: VIPER targets PIM designs before they exist.
% The design points in Tables~\ref{tab:task_off_configs}
% and~\ref{tab:val_configs} span SRAM, DRAM, HBM, NVM, and
% storage-class PIM at arbitrary latencies and capacities, and no
% silicon implementation of this space exists to measure against.
% Cycle-accurate full-system simulation is therefore the only
% available ground truth, and gem5 is the
% pre-silicon reference, itself extensively validated against real
% hardware \cite{qureshi2019gem5x, gem5_accuracy}. To further reduce dependence
% on absolute simulator fidelity, our data triggering evaluation
% compares normalized performance trends rather than raw execution
% times (Section~\ref{sec:val_data}). The UPMEM case study
% (Section~\ref{sec:upmem_case}) complements this validation with a
% hardware anchor, since its inputs are measured host performance
% counters and published constants from a commercially shipping PIM
% device.

\subsection{Task Offloading Validation}
\label{sec:val_task}

To validate the task offloading model, we use gem5-\-SALAM \cite{ceary2020gem5salam}, a full-system extension of gem5 that models hardware accelerators at the LLVM-IR level alongside a gem5 CPU and memory hierarchy. gem5-SALAM exposes per-LLVM-instruction latency through a YAML configuration, which lets us sweep accelerator compute and memory parameters without modifying the accelerator model itself. The simulated platform is a single in-order ARM \texttt{MinorCPU} with the default L1/L2 cache hierarchy and an 8~GB DDR4-2400 main memory. The accelerator under test is a GEMM cluster composed of a top controller, a compute engine modeled from the LLVM-IR of a GEMM kernel, three operand scratchpads (\texttt{MATRIX1/2/3}), and a non-coherent DMA engine that streams operands between DDR and the scratchpads. By varying the scratchpad access latency $m$, this cluster can represent both PIM paradigms: a high $m$ models a PNM accelerator where every dispatched tile must first move operand data from the memory array into the scratchpads ($t_{array}$) before compute begins, while $m{=}0$ models a PUM accelerator where operands reside in place and no array-to-PE transfer is required.

The benchmark is a single-layer Llama2 transformer prefill pass with $\text{dim}=128$, $\text{hidden\_dim}=256$, 8 heads, vocabulary size 256, and a batch of $B=128$ tokens per forward pass. The pipeline runs the standard Llama2 stack (embedding, RMSNorm, RoPE, multi-head attention, SwiGLU FFN, and classifier head), and every matmul is dispatched to the accelerator as a sequence of tile-sized sub-matmuls. For each sub-matmul the CPU issues a DMA into the scratchpads, triggers the compute kernel, waits for completion, and DMAs the result back. This dispatch loop reproduces the per-instruction array-to-PE transfer that characterizes PNM task offloading.

\begin{table}[t]
\centering
\caption{Task offloading validation parameter sweep.}
\vspace*{-3mm}
\label{tab:task_off_configs}
\small
\begin{tabular}{ll}
  
\toprule
\textbf{Parameter} & \textbf{Values} \\
\midrule
GEMM tile size (ROW=COL) & 8, 16, 32, 64, 128, 256 \\
PIM compute latency $c$ (cycles) & 5, 9, 14, 18, 50, 500, 5000 \\
Scratchpad latency $m$ (ns) & 0, 2, 4, 8, 16, 32, 64, 128 \\
\bottomrule
\end{tabular}
\vspace*{-3mm}
\end{table}

We sweep three accelerator design knobs, summarized in Table~\ref{tab:task_off_configs}: the GEMM tile size, the per-FPU compute latency $c$ (where $c \in \{5,9,14,18\}$ correspond to realistic pipelined-FPU MAC depths and $c \in \{50, 500, 5000\}$ are stress values that push the accelerator into a compute-bound regime), and the scratchpad access latency $m$ ($m=0$: PUM, no array-to-PE transfer; $m>0$: PNM with increasing data-movement cost). The full grid covers $6 \times 7 \times 8 = 336$ configurations.

\begin{figure}
  \centering
  \resizebox{1.00\columnwidth}{!}{\includegraphics{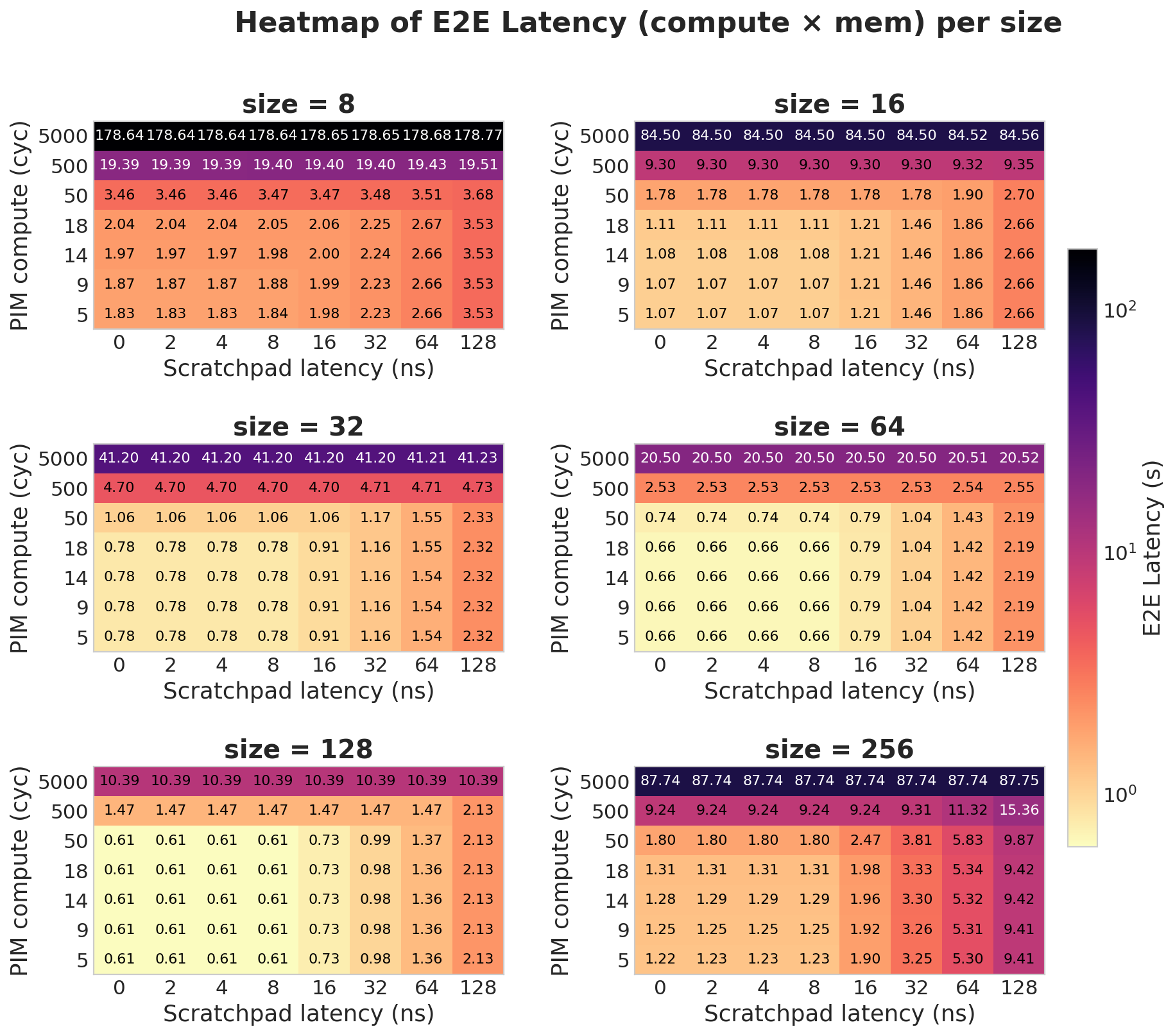}} % Adjust the scale as needed
  \vspace*{-6mm}
  \caption{gem5-SALAM end-to-end latency across the 336-point sweep by tile size. Memory-bound regime (low $c$): runtime scales with $m$; compute-bound regime (high $c$): $m$ has no effect.}
  \label{fig:task_offloading}
  \vspace*{-3mm}
\end{figure}

\begin{figure}
  \centering
  \resizebox{0.95\columnwidth}{!}{\includegraphics{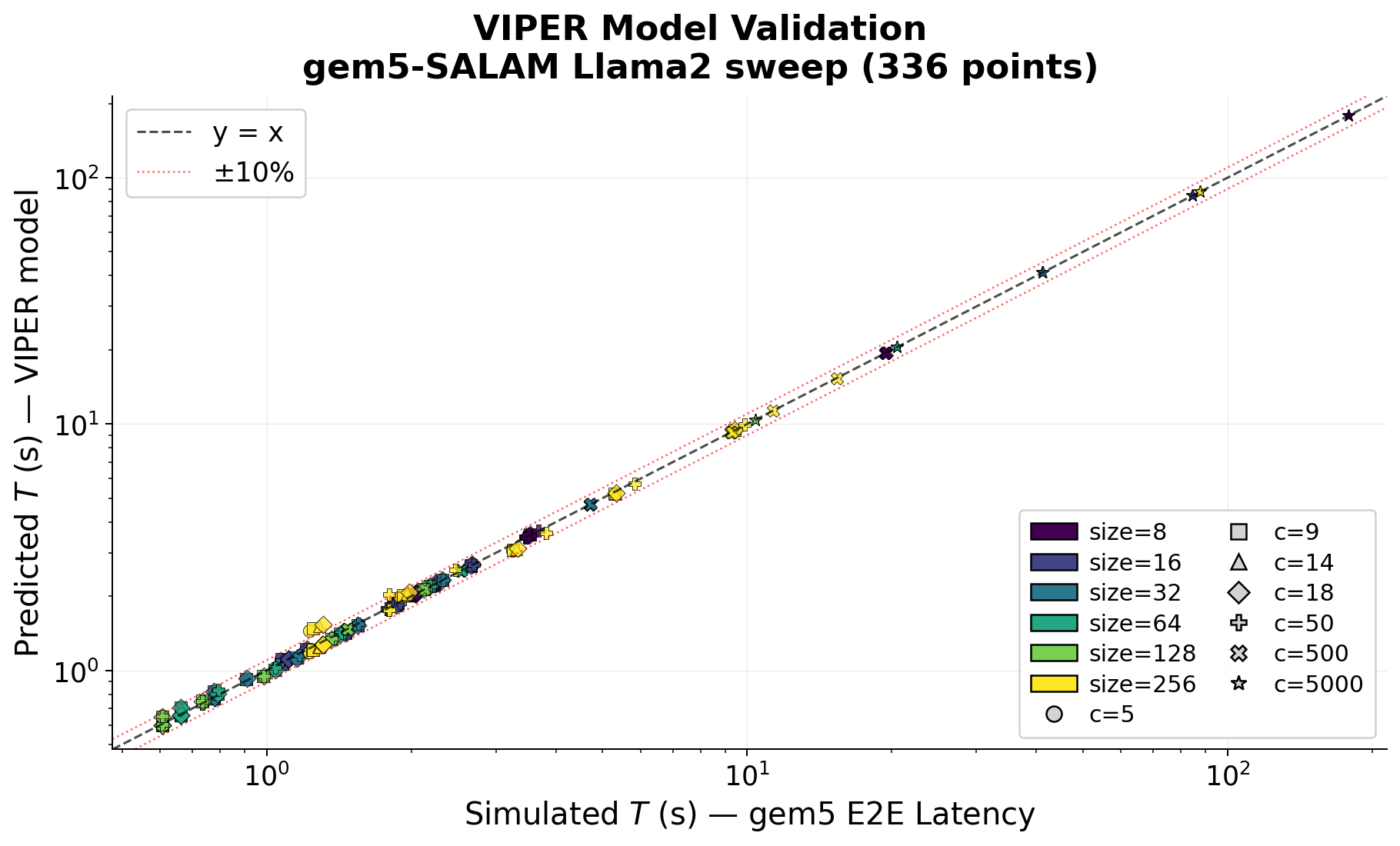}}
  \vspace*{-3mm}
  \caption{Predicted vs.\ simulated end-to-end latency (log--log, 336 points). Dotted red band: $\pm 10\%$ envelope.}
  \label{fig:viper_scatter}
  \vspace*{-2mm}
\end{figure}

Figure~\ref{fig:task_offloading} shows the measured end-to-end latency across the full sweep, broken down per tile size. The heatmaps confirm the two regimes that the VIPER task offloading model predicts. In the compute-bound regime ($c=500, 5000$), runtime is essentially flat across $m$, the accelerator spends all its time in the compute engine and the scratchpad latency is fully hidden, matching VIPER's prediction that $\text{CPI}_{PIM,exec}$ dominates. In the memory-bound regime ($c \leq 50$), runtime grows monotonically with $m$, with the largest tiles (size 64, 128, 256) showing up to a $3.5\times$ slowdown between $m=0$ and $m=128$. The quantitative agreement between VIPER's predictions and the simulated runtimes is established in Figure~\ref{fig:viper_scatter}.

% VIPER reproduces the observed transition point and the slope of the memory-bound region across all six tile sizes, demonstrating that the analytical task offloading model captures both the compute- and memory-bound behavior of a real PNM accelerator without requiring cycle-accurate simulation.

Figure~\ref{fig:viper_scatter} provides a predicted-vs-simulated scatter over the entire sweep. The points sit tightly along the $y=x$ diagonal across all three decades of execution time, with $\sim 326$ of $336$ points (97.0\%) falling inside the $\pm 10\%$ envelope. The analytical model captures scaling with tile size $T$: neither the small-time cluster (size 8, low $c$) nor the large-time cluster (size 256, high $c$) drifts off the diagonal as a group, so there is no size- or compute-dependent bias the model fails to capture.

\subsection{Data Triggering Validation}
\label{sec:val_data}

To validate the data triggering model, we extend gem5-\- SALAM~\cite{ceary2020gem5salam} to model PIM behavior directly within the memory hierarchy. Specifically, we modify the cache and memory controller to inject a configurable computation delay when a data triggering event occurs at a given memory level. When the PIM device has sufficient capacity, it processes the data locally. When capacity is exceeded, the PIM device falls through to the DRAM to fetch the remaining data. We compare the gem5 execution results against VIPER predictions, where PIM-side parameters are set to match the gem5 configuration and CPU-side parameters are collected from our server using Linux \texttt{perf}. For simplicity, we use a single aggregate computation latency $t_{PIM,\text{mem}}$ without decomposing it into $t_{array}$ and $t_{prog}$, as the goal here is to validate the architecture-aware timing model.

\begin{table}[t]
\centering
\caption{Data triggering validation configurations.}
\vspace*{-3mm}
\label{tab:val_configs}
\small
\begin{tabular}{lllll}
\toprule
\textbf{ID} & \textbf{Configuration} & \textbf{CPU} & \textbf{Trigger Level} & \textbf{Represents} \\
\midrule
C1  & 5ns, 1MB    & In-order & After L1  & SRAM PNM \\
C2  & 5ns, 1MB    & OoO      & After L1  & SRAM PNM \\
C3  & 50ns, 1MB   & In-order & After LLC & DRAM PNM \\
C4  & 50ns, 1MB   & OoO      & After LLC & DRAM PNM \\
C5  & 500ns, 1MB  & In-order & In DRAM   & NVM PUM \\
C6  & 500ns, 1MB  & OoO      & In DRAM   & NVM PUM \\
C7  & 50ns, 64MB  & In-order      & After LLC & SRAM PUM  \\
C8  & 50ns, 256MB & In-order      & After LLC & HBM PNM \\
C9  & 50ns, 4GB   & In-order      & After LLC & DRAM  PNM \\
C10 & 500ns, 100GB & In-order & In DRAM  & Storage PNM \\
C11 & 500ns, 100GB & OoO     & In DRAM   & Storage PNM \\
\bottomrule
\vspace*{-8mm}
\end{tabular}
\end{table}

We evaluate across 11 configurations representing a broad range of PIM design points, as summarized in Table~\ref{tab:val_configs}, using 12 benchmarks drawn from SPEC2017 \cite{spec2017} CPU and GraphBig \cite{graphbig} with a GitHub developer social network dataset \cite{graph_data_set}.  We test both in-order and out-of-order (OoO) CPUs since VIPER does not model memory-level parallelism~(MLP). In-order CPUs stall sequentially, so VIPER predictions are expected to be highly accurate. OoO CPUs overlap multiple memory requests, which VIPER cannot capture, so we expect and quantify higher prediction error under OoO execution.

\begin{figure}
  \centering
% \vspace*{-1mm}
 \resizebox{1.00\columnwidth}{!}{\includegraphics{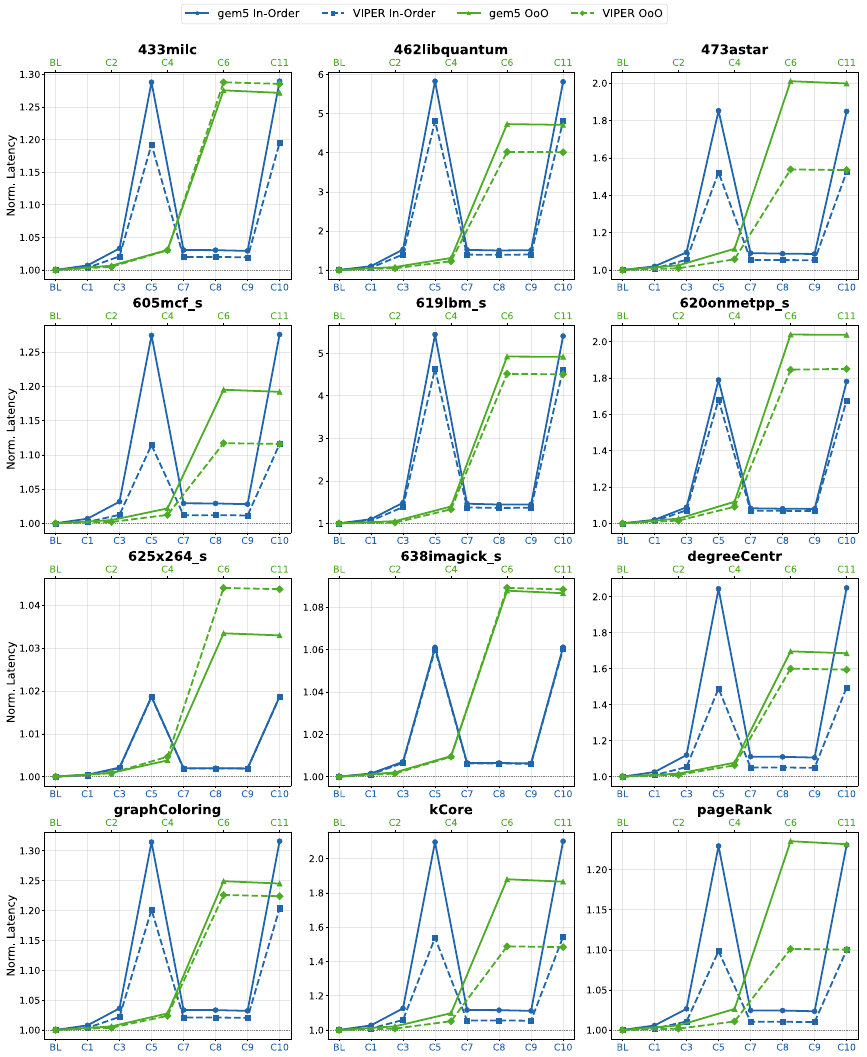}} % Adjust the scale as needed
 \vspace*{-5mm}
  \caption{Normalized latency (relative to no-PIM baseline) across all 11 configurations (Table~\ref{tab:val_configs}) and 12 benchmarks. Solid: gem5; dashed: VIPER.}
  \label{fig:Trend}
  % \vspace*{-5mm}
\end{figure}

To evaluate VIPER’s accuracy, we compare the normalized latency trend of each benchmark under PIM against the no-PIM baseline. Directly comparing execution times would be misleading because gem5 and VIPER operate at fundamentally different abstraction levels and use different ISAs. Therefore, we focus on relative performance changes and evaluate whether VIPER correctly captures the trends introduced by PIM.

Figure~\ref{fig:Trend} shows the normalized latency trend across all 11 configurations for each benchmark. Overall, VIPER closely tracks the gem5 trend across both SPEC and GraphBig benchmarks. For in-order CPU results, VIPER and gem5 follow similar trends across all configurations. The dashed and solid lines are visually indistinguishable for most SPEC benchmarks, such as \texttt{625.x264\_s} and \texttt{638.imagick\_s}, which are compute-bound and exhibit low memory sensitivity. Memory-bound SPEC benchmarks such as \texttt{462.lib\-quantum} and \texttt{619.lbm\_s} show larger normalized latencies but still track the trend well. GraphBig benchmarks show a slightly larger divergence between VIPER and gem5, particularly under OoO execution, due to their highly irregular memory access patterns that exhibit stronger MLP effects. For OoO results, VIPER consistently overestimates the normalized latency relative to gem5, as expected since the analytical model does not account for memory request overlapping. Despite this, the trend direction is preserved in all cases.

\begin{figure}
  \centering
% \vspace*{-1mm}
 \resizebox{1.00\columnwidth}{!}{\includegraphics{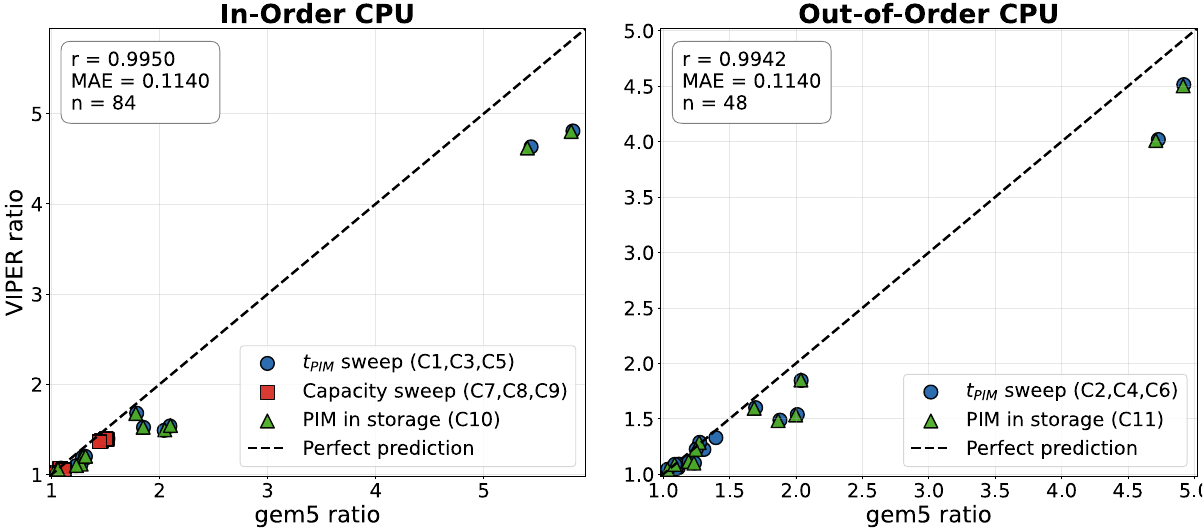}} % Adjust the scale as needed
 \vspace*{-5mm}
 \caption{VIPER vs.\ gem5 normalized latency
 ($\text{latency}_{PIM}/\text{latency}_{baseline}$) for all
 benchmark--configuration pairs (Table~\ref{tab:val_configs}).
 Diagonal: perfect agreement. Each panel reports Pearson $r$, mean
 absolute error, and point count $n$ over its own configuration set;
 the near-identical MAEs ($0.11404$ vs.\ $0.11397$) are coincidental.}
  \label{fig:Scattered}
  \vspace*{-5mm}
\end{figure}

Figure~\ref{fig:Scattered} provides a quantitative view of VIPER prediction accuracy. For in-order CPU, VIPER achieves a Pearson correlation of $r = 0.995$ and a mean absolute error of $\text{MAE} = 0.114$, which corresponds to 2.4\% of the full ratio range. Points from the $t_{PIM}$ sweep and capacity sweep configurations cluster tightly along the diagonal, indicating consistent accuracy across both timing and capacity dimensions. Notably, VIPER correctly predicts the capacity wall behavior across C7, C8, and C9, confirming that the partitioning model accurately captures the performance impact of exceeding device capacity. For OoO CPU, the Pearson correlation remains high at $r = 0.994$ with $\text{MAE} = 0.114$, but points shift above the diagonal, reflecting systematic overestimation caused by unmodeled MLP. The most notable outliers correspond to GraphBig benchmarks under high-latency configurations (C5, C6, C10, C11), where irregular memory access patterns amplify the modeling error. Overall, VIPER achieves a mean prediction error of 4.4\% on in-order regular memory access (SPEC), 7.3\% on in-order irregular access (GraphBig), 4.9\% on OoO regular access, and 5.5\% on OoO irregular access, demonstrating reliable trend prediction across all evaluated configurations.

In summary, VIPER tracks gem5’s normalized-latency trends while completing the analysis in under one minute, versus 6-10
hours per benchmark in gem5

\section{Real PIM Study}
\label{sec:Result}

We apply VIPER to UPMEM~\cite{devaux2019upmem, gomez2021benchmarking}, ReRAM/FeFET crossbars~\cite{xx_nvsim}, and IMCRYPTO~\cite{imcrypto} to study task-offloading and data-triggered PIM designs

% We apply VIPER to three real PIM designs.  This section demonstrates why a designer needs them. Each case study asks a design question that circuit- and device-level tools cannot answer, and shows the insight VIPER surfaces in seconds. We study UPMEM~\cite{devaux2019upmem, gomez2021benchmarking}as a PNM representative, and ReRAM/FeFET crossbars~\cite{xx_nvsim} and IMCRYPTO~\cite{imcrypto} as PUM representatives under task offloading and data triggering respectively.

% We apply VIPER to two real real PIM designs to demonstrate how a designer can use VIPER for early-stage DSE and system-level performance estimation. We study IMCRYPTO\cite{imcrypto} as a PUM representative and UPMEM~\cite{devaux2019upmem, gomez2021benchmarking} as a PNM representative.
\subsection{PNM Case Study: UPMEM for LLM Inference}
\label{sec:upmem_case}

UPMEM~\cite{devaux2019upmem, gomez2021benchmarking} is the first commercially available PNM system, integrating 2{,}560 in-order DPUs at 350~MHz directly inside DRAM ranks. Each DPU has a small WRAM scratchpad backed by its local MRAM bank. We model UPMEM as a task-offloading PNM accelerator and use VIPER in two stages: we first predict the end-to-end speedup of a single-layer Llama2 prefill ($\text{dim}=128$, $\text{hidden\_dim}=256$, 8 heads, $B=128$) when all eight matmuls per forward pass are offloaded to the DPUs, \emph{without running the workload on UPMEM hardware}, and we then validate the prediction on a real 2{,}560-DPU UPMEM server. The a-priori VIPER inputs come from two independent sources: measured CPU perf counters for the non-offloaded term, and published UPMEM constants for the offloaded term.

\vspace{2pt}\noindent\textbf{A priori prediction.} To isolate the
host work we build the forward pass twice in Q16.16: a full version
and a stub that replaces \texttt{matmul\_batch} with a same-shape
\texttt{memset} (RMSNorm, RoPE, softmax, SwiGLU remain). We measure
$T_{\text{full}}=20.37$ and $T_{\text{stub}}=3.10$~ms, so the matmuls
are 84.8\% of runtime and $T_{\text{full}}/T_{\text{stub}}=6.57\times$
is the Amdahl ceiling on matmul offload.

The offloaded term uses published UPMEM constants: per-DPU INT32
throughput of 45~MOPS~\cite{upmem_unleashed},
CPU$\leftrightarrow$DPU bandwidths of 2.5/2.0~GB/s~\cite{gomez2021benchmarking},
and 0.1~ms per-call launch overhead~\cite{pgemmlib_cgo25}.  Across the
eight matmuls the workload performs 25.2~M MACs and transfers 1.28~MB
of activations.  With weights resident in MRAM, $t_{array}$ collapses
to a local access, so the PIM memory term of
Equation~\ref{eq:pnm_task_offload} reduces to the
host$\leftrightarrow$DPU transfers $t_{in}+t_{out}$ plus launch
overhead, about 1.40~ms total.  The per-DPU kernel time is
$T_{\text{kernel}}(1)=25.17\,\text{M}/45\,\text{MOPS}=559.24$~ms.

In the task-offloading model (Equation~\ref{eq:task_offload}), the CPU
term is $T_{\text{stub}}=3.10$~ms and the PIM term combines the
parallelized kernel $T_{\text{kernel}}(1)/N$ with the $1.40$~ms
transfer-plus-launch cost; the working set fits in MRAM, so
$m_{PIM}=1$.  The prediction is
$T_{\text{total}}(N)=3.10+559.24/N+1.40$~ms.
Sweeping $N$, VIPER predicts break-even at $N\approx35$ DPUs and a
speedup climbing monotonically toward ${\sim}4.5\times$ (reaching
$4.3\times$ on the full 2{,}560-DPU server), held below the
$6.57\times$ Amdahl ceiling only by the fixed 1.40~ms transfer floor.

% Before porting a line of code the decision is already clear: offloading
% is worth it, a few dozen DPUs pay for themselves, and the achievable
% prize is several-fold.

\vspace{2pt}\noindent\textbf{Validation on real hardware.}
We implement the offload on a 2{,}560-DPU UPMEM server (350~MHz,
SDK~2025.1.0, Intel Xeon Silver 4215 host, the same CPU used for the
a-priori profile; best-of-5 with one warmup).  The eight matmuls run as
INT8 GEMM tiles with INT32 accumulation, weights resident in MRAM and
fused into five dispatches per pass; output matches the
Q16.16 reference to 3.3\% relative L2 error.

Hardware confirms the compute side: measured per-DPU throughput is
48.0~MOPS, within 7\% of the assumed 45; the kernel follows
$T_{\text{kernel}}(1)/N$ to within ${\sim}2\%$ up to 512 DPUs; and
break-even lands at $N\approx40$--$49$ DPUs, close to the predicted 35.
The measured break-even range and speedup bound confirm the a priori offload decision.

% The go/no-go call and the Amdahl ceiling held before any
% code was ported.

\begin{figure}
  \centering
 \resizebox{0.99\columnwidth}{!}{\includegraphics{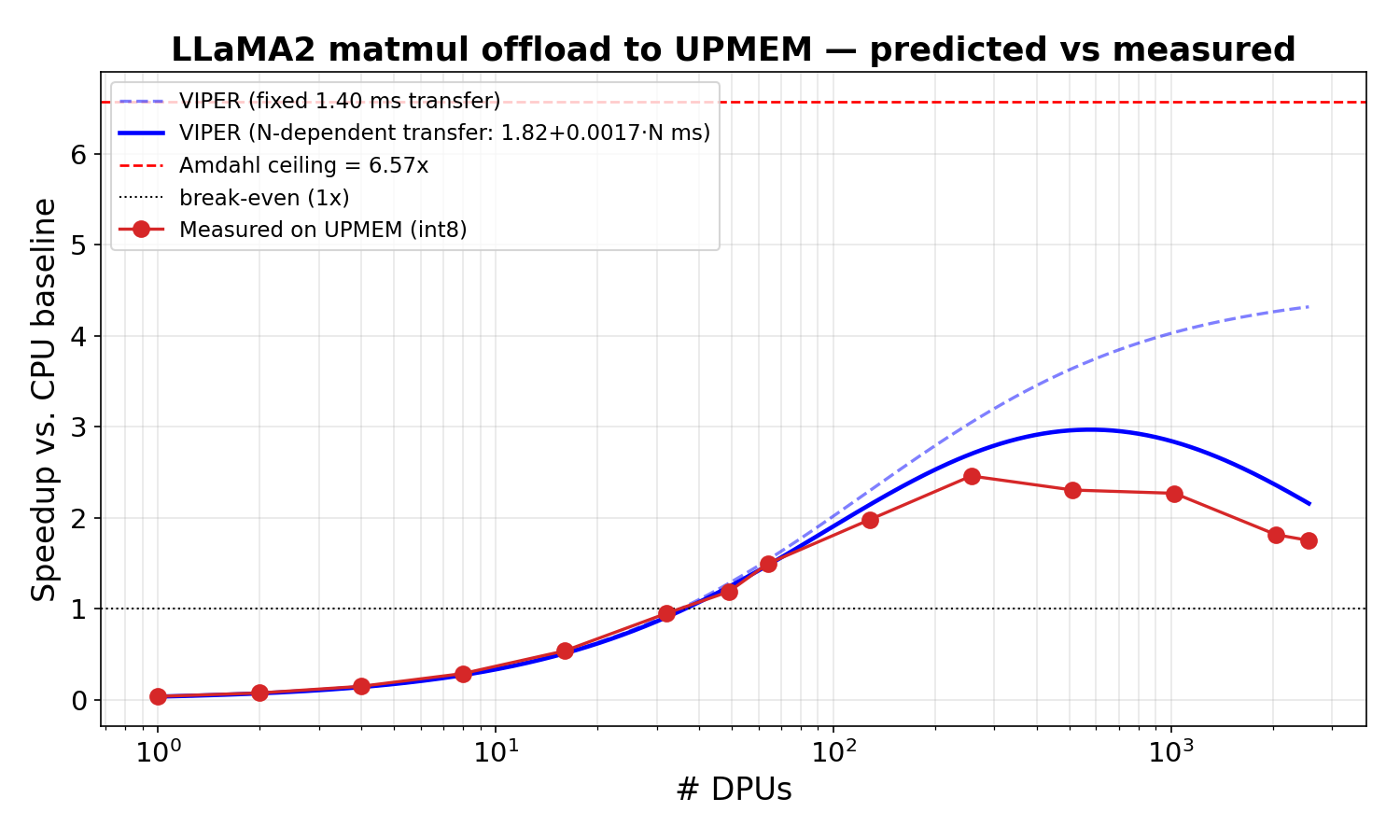}}
 % \vspace*{-5mm}
  \caption{Predicted vs.\ measured UPMEM speedup for single-layer
  Llama2 matmul offload.  The fixed-transfer prediction climbs
  monotonically, but both refined VIPER and hardware peak near 256 DPUs
  and roll off as host-mediated transfer grows with DPU count.}
  \label{fig:upmem_speedup}
  \vspace*{-5mm}
\end{figure}

The a-priori miss is confined to a single term.  VIPER predicted
$4.3\times$ on the full server; instead the speedup peaks at
$2.46\times$ at 256 DPUs and then \emph{declines}, yielding only
$1.75\times$ at 2{,}560 DPUs (Figure~\ref{fig:upmem_speedup}).  The
cause is the aggregate transfer $t_{in}+t_{out}$, which the a-priori
model held constant at 1.40~ms: the measured value stays near it below
${\sim}64$ DPUs
but grows to ${\sim}5.9$~ms at 2{,}560 DPUs, because with no inter-DPU
communication path every activation scatter and partial-sum gather
routes through the host over a shared transfer path, and each additional DPU increases descriptor-setup, transfer-programming, and synchronization overhead.  Refining that
one input, letting the aggregate $t_{in}+t_{out}$ grow with DPU count
as $1.82+0.00168\,N$~ms, turns the monotone climb into a
peak-and-rolloff: the workload stays host-work-bound at moderate DPU
counts, as predicted, but becomes transfer-bound at high counts, so an
\emph{optimal} DPU count replaces monotonic saturation.  A single measured input is enough to bring VIPER into agreement with
hardware: swapping the constant transfer for the affine $t_{in}+t_{out}$
reproduces the peak-and-rolloff the fixed model cannot, holds the
prediction within the same $2$--$3\times$ band as the measurements, and
peaks at ${\sim}3.0\times$ near ${\sim}570$ DPUs against the measured
$2.46\times$ at 256 (Figure~\ref{fig:upmem_speedup}).  More importantly,
VIPER now exposes the design-critical fact no datasheet reveals: past a
few hundred DPUs, adding more \emph{lowers} end-to-end speedup.  The remaining error arises because the affine fit slightly understates
the transfer at intermediate $N$, and the kernel departs from $1/N$
scaling past 512 DPUs.

% In summary, the a-priori model got the design \emph{decision} right from
% independent inputs (the $6.57\times$ Amdahl ceiling, the $1/N$ compute
% scaling, and break-even at a few dozen DPUs) but overpredicted the
% achievable peak because it assumed a flat transfer term.  Refining that
% one input flips the predicted trend from monotonic saturation to a
% peaked curve in the same $2$--$3\times$ band as the hardware, correctly
% signaling that a DPU sweet spot exists (measured at 256).
\begin{center}
\setlength{\fboxrule}{1.5pt}
\fcolorbox{red!60!black}{red!10}{%
\parbox{0.95\linewidth}{%
\textcolor{red!60!black}{%
\textbf{Observation from UPMEM case study:} For PNM accelerators that offload one operator class, VIPER exposes the Amdahl ceiling set by non-offloadable host work \emph{before} any code is ported. Hardware then reveals the binding limit is tighter still: host-mediated transfer and per-dispatch overhead cap the achievable speedup well below that ceiling and create an \emph{optimal} DPU count, rather than monotonic saturation. VIPER captures this limit by refining a single transfer input.}
}}
\end{center}

% \begin{center}
% \setlength{\fboxrule}{1.5pt}
% \fcolorbox{red!60!black}{red!10}{%
% \parbox{0.95\linewidth}{%
% \textcolor{red!60!black}{%
% \textbf{Observation from UPMEM case study:} For PNM accelerators
% that offload one operator class, VIPER exposes the design
% decision \emph{before} any code is ported: from independent
% inputs alone, the a-priori model correctly predicted the
% $6.57\times$ Amdahl ceiling set by non-offloadable host work, the
% $1/N$ compute scaling, and break-even at a few dozen DPUs.
% Hardware then reveals the binding limit is tighter still:
% host-mediated transfer and per-dispatch overhead cap the
% achievable speedup well below that ceiling and create an
% \emph{optimal} DPU count rather than monotonic saturation.
% Refining that single transfer input flips the predicted trend
% from monotone climb to a peaked curve in the same
% $2$--$3\times$ band as the hardware, correctly signaling the DPU
% sweet spot (measured at 256).}
% }}
% \end{center}

\subsection{PUM Case Study: XBAs with a Circuit-Accurate Simulator}

\begin{figure}
  \centering
 \resizebox{1.00\columnwidth}{!}{\includegraphics{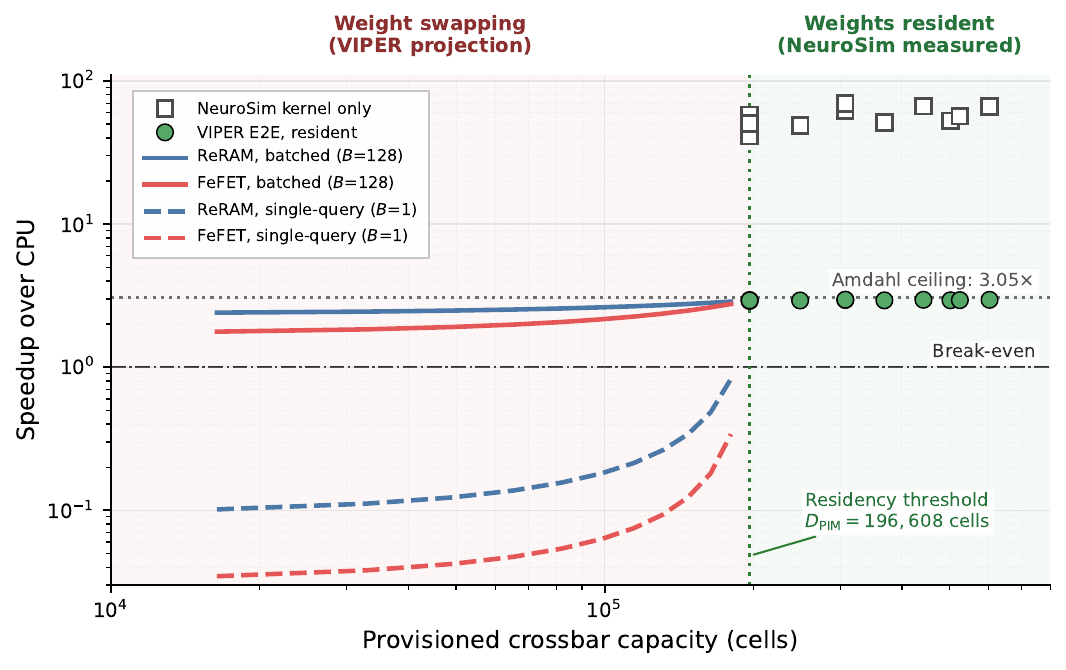}}
 \vspace*{-5mm}
  \caption{XBA case study for an eight-layer DNN. Squares and circles show kernel-only and VIPER end-to-end speedups for resident weights, respectively.}
  \label{fig:xba_case}
  \vspace*{-5mm}
\end{figure}

Crossbar arrays (XBAs) are the flagship PUM substrate: an
$N{\times}N$ array of resistive cells stores a weight tile in cell
conductances and performs a matrix--vector multiplication in a
single $O(1)$ analog read~\cite{shafiee2016isaac, chi2016prime}.
Weights live in device state, and NVM programming is far slower
than reading~\cite{XX_reram_write, ni2019fefet_tcam}, so any weight
that does not fit in provisioned capacity must be reprogrammed on
the critical path.  \textbf{ We use VIPER to quantify how host execution, capacity limits, and programming latency change circuit-level
speedup estimates.}

The workload offloads eight DNN weight matmuls
($D_{PIM}=196{,}608$ cells); the host baseline is profiled on our
desktop CPU with a stub binary that removes the matmuls
($T_{\text{full}}{=}18.92$\,ms, $T_{\text{stub}}{=}6.21$\,ms,
$B{=}128$), bounding any offload by a $3.05\times$ Amdahl ceiling.
We run DNN+NeuroSim~\cite{chen2018neurosim} (8-bit, 22\,nm) on this
network for every subarray size it supports (32--256): NeuroSim
instantiates as many arrays as the tiled mapping requires, so each
run is a complete resident chip, e.g., 192 arrays of $32^2$
(2.41\,\textmu s per inference), 12 of $128^2$ (1.97\,\textmu s),
or 8 of $256^2$ (1.75\,\textmu s). In
Fig.~\ref{fig:xba_case}, each such chip is placed at its total
capacity, arrays${\times}N^2$ cells (x-axis), against speedup over
the CPU (y-axis, log): squares divide the CPU matmul time by
NeuroSim's latency alone (kernel view), circles feed the same
latency into Equation~\ref{eq:task_offload} as $t_{comp}$ with the profiled host term.

% \begin{center}
% \setlength{\fboxrule}{1.5pt}
% \fcolorbox{red!60!black}{red!10}{%
% \parbox{0.95\linewidth}{%
% \textcolor{red!60!black}{%
% \textbf{Observation from XBAs case study:} VIPER agrees with the
% circuit-accurate simulator inside its regime and extends beyond
% it. A $41$--$70\times$ kernel never delivers more than
% $2.96\times$ end to end, because host work dominates once the
% accelerator is fast. Below the residency threshold, the missing
% tiles must be reprogrammed on every pass through the weights:
% batched inference ($B{=}128$) amortizes each pass and retains
% $1.8$--$2.9\times$, while single-query serving makes a full pass
% per query and collapses to $0.04$--$0.1\times$. And the two
% technologies, byte-identical in NeuroSim's output because its
% inference estimator assigns no write latency, separate by
% $3\times$ the moment programming enters the critical path. None
% of this is visible from a circuit-accurate inference simulator
% alone.}
% }}
% \end{center}

The gap between the two is the first result: a
$41$--$70\times$ kernel never delivers more than $2.96\times$ end
to end, because host work dominates once the accelerator is fast.
Left of the threshold, VIPER models what NeuroSim structurally
cannot: with only $A$ of the 12 arrays provisioned, the missing
tiles must be reprogrammed on every pass through the weights ($N$
rows at a write-verified pulse of 0.1/0.3\,\textmu s for
ReRAM/FeFET~\cite{XX_reram_write, ni2019fefet_tcam}); batched
inference shares each pass across $B{=}128$ queries and retains
$1.8$--$2.9\times$, while single-query inference makes a full pass
per query and collapses to $0.04$--$0.1\times$.  NeuroSim assigns identical latency to the two technologies because it omits write latency; VIPER predicts a 3$\times$
difference once programming is included..

\begin{center}
\setlength{\fboxrule}{1.5pt}
\fcolorbox{red!60!black}{red!10}{%
\parbox{0.95\linewidth}{%
\textcolor{red!60!black}{%
\textbf{Observation from XBAs case study:}  For resident weights, VIPER agrees with the circuit-level estimate and extends the analysis to capacity-limited
configurations. A $41$--$70\times$ kernel yields at most $2.96\times$ end to end;
below the residency threshold, single-query serving collapses while
batching survives; and the ReRAM/FeFET choice matters only where
programming enters the critical path. None of this is visible from
a circuit-accurate inference simulator alone.}
}}
\end{center}

\subsection{PUM Case Study: IMCRYPTO in Secure Memory}
IMCRYPTO~\cite{imcrypto} implements the full AES-128 pipeline inside
SRAM arrays: post-layout results at 22\,nm show encryption in 30
cycles at 1\,GHz on a 3\,KB dual-port array. We model it as a
data-triggered PUM gateway on the CPU--DRAM path, the one case study
exercising VIPER's data-triggering model (Equations~\ref{eq:amat_pim}): unlike
offloading, a triggered design lies on the critical memory path and adds latency to every DRAM access. We ask: \textbf{what does
in-memory encryption cost the memory system, and which device
technologies can afford to host it?}

We model IMCRYPTO as a data-triggered PUM gateway on the
CPU--DRAM path, such that every DRAM access reaching the gateway
incurs in-memory AES encryption or decryption. For each
SPEC2017~\cite{spec2017} and GraphBig~\cite{graphbig} benchmark,
we profile an Intel Xeon E-2388G using Linux \texttt{perf} to
obtain the cache miss rates ($m_{L1}$, $m_{L2}$, $m_{L3}$) and
construct the unsecured baseline AMAT from the corresponding
cache and DRAM latencies. The IMCRYPTO
timing parameters come from post-layout and gem5 results:
$t_{in}=t_{out}=10$, $t_{comp}=30$, and $t_{prog}=1$\,ns for
data transfer, AES-128 computation, and SRAM programming, giving
$t_{PIM,mem}=t_{in}+t_{prog}+t_{comp}+t_{out}=51$\,ns. VIPER
substitutes this latency for $t_{DRAM}$ in
Equation~\ref{eq:amat_pim} and normalizes the resulting AMAT to
the unsecured DRAM baseline. Across all 12 benchmarks and gateway
capacities of 64\,MB, 256\,MB, and 4\,GB, VIPER reproduces the
gem5-measured normalized AMAT to within $0.2\%$. The limited
sensitivity to capacity indicates that the overhead is dominated
by data transfer and computation rather than capacity constraints.
Overall, securing all DRAM traffic increases AMAT by $1.5\%$ on
geometric mean and by $6.2\%$ in the worst case
(\texttt{462.libquantum}).

\begin{figure}
  \centering
% \vspace*{-1mm}
 \resizebox{1.00\columnwidth}{!}{\includegraphics{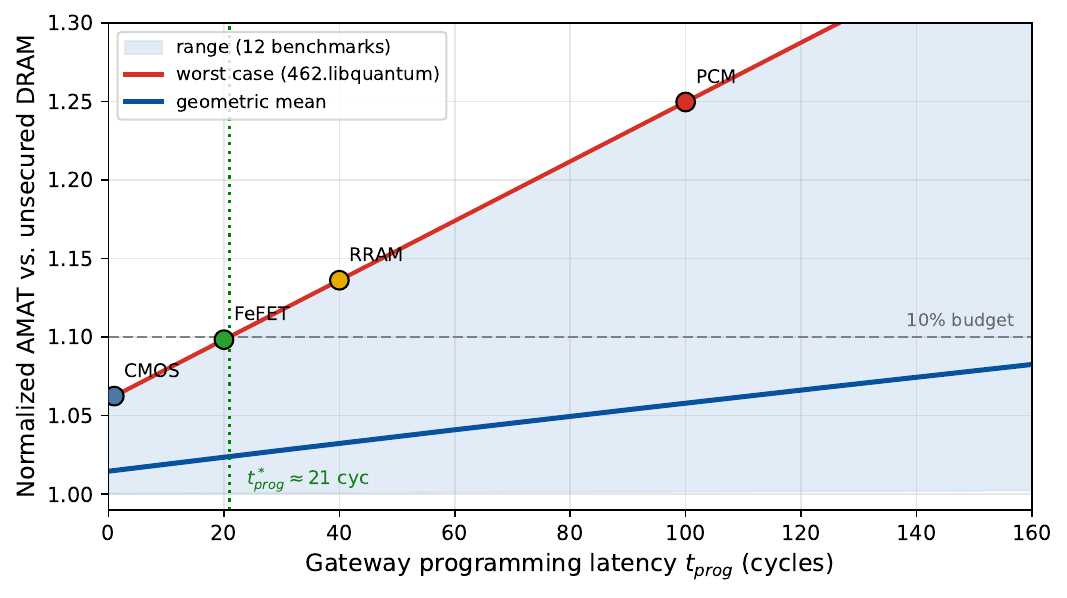}} % Adjust the scale as needed
 \vspace*{-8mm}
  \caption{Normalized AMAT vs.\ gateway programming latency
$t_{prog}$ on measured host profiles; CMOS, FeFET, RRAM, and PCM sit
at their respective values, and a $10\%$ worst-case budget admits
only $t_{prog} \lesssim 21$ cycles. }
  \label{fig:imcrypto_tprog}
  \vspace*{-5mm}
\end{figure}

% \begin{center}
% \setlength{\fboxrule}{1.5pt}
% \fcolorbox{red!60!black}{red!10}{%
% \parbox{0.95\linewidth}{%
% \textcolor{red!60!black}{%
% \textbf{Observation from IMCRYPTO case study:} A data-triggered
% PIM lives on the memory critical path, so its cost is paid on
% every access. Measured on real host profiles, always-on
% in-memory encryption costs only $1.5\%$ mean AMAT, and VIPER
% matches the gem5-measured cost within $0.2\%$. Sweeping
% $t_{prog}$ then answers what post-layout numbers cannot: holding
% the worst-case tax under a $10\%$ budget requires
% $t_{prog} \lesssim 21$ cycles, so CMOS ($6.2\%$) and FeFET
% ($9.8\%$) qualify as gateway substrates while RRAM ($13.6\%$)
% and PCM ($25.0\%$) do not, despite their density and
% non-volatility advantages. Technology selection for a
% critical-path PIM reduces to a single quantitative test.}
% }}
% \end{center}

VIPER then answers what the post-layout numbers cannot:
Fig.~\ref{fig:imcrypto_tprog} sweeps the array programming latency
$t_{prog}$, placing CMOS, FeFET, RRAM, and PCM at their respective
values. Keeping worst-case overhead below $10\%$ budget requires
$t_{prog} \lesssim 21$ cycles: CMOS ($6.2\%$) and FeFET ($9.8\%$)
qualify as gateway substrates, while RRAM ($13.6\%$) and PCM
($25.0\%$) do not, despite their density and non-volatility
advantages. This sweep provides a quantitative screening criterion for critical-path PIM technologies.

\begin{center}
\setlength{\fboxrule}{1.5pt}
\fcolorbox{red!60!black}{red!10}{%
\parbox{0.95\linewidth}{%
\textcolor{red!60!black}{%
\textbf{Observation from IMCRYPTO case study:}  A data-triggered PIM lies on the memory critical path and adds latency to every access. Measured on real profiles, always-on in-memory encryption
costs only $1.5\%$ mean AMAT, and VIPER matches the gem5-measured
cost within $0.2\%$; sweeping $t_{prog}$ then shows only CMOS and
FeFET stay within a $10\%$ worst-case budget, a substrate decision
invisible from post-layout numbers alone.}
}}
\end{center}

\section{Conclusion}
\label{sec:conclusion}
We present VIPER, a lightweight, cross-layer analytical framework
for PNM and PUM design-space exploration, achieving below 10\% mean
error across more than 400 cycle-accurate gem5 configurations.
Measurements on a 2,560-DPU UPMEM system further validate the
predicted break-even behavior and end-to-end bottlenecks. Case
studies on real designs show that cross-layer costs invisible
at the circuit level, such as programming latency, capacity
limits, and non-offloadable host work, fundamentally
determine whether a PIM design delivers real benefit. This work
addresses performance; extending the formulation with per-access
energy terms for joint performance and energy DSE is future
work. VIPER is open-sourced at [URL redacted for blind review].

\bibliographystyle{IEEEtranS}
\bibliography{refs}

\end{document}